\documentclass[aps,prl,twocolumn,amsmath,amssymb,showpacs,superscriptaddress,notitlepage,longbibliography,floatfix]{revtex4-2}
\usepackage{graphicx}
\usepackage{subfigure}
\usepackage{dcolumn}
\usepackage{bm}
\usepackage{amsfonts}
\usepackage{mathrsfs}
\usepackage{amssymb}
\usepackage{amsmath}
\usepackage{color}
\usepackage{chngcntr}
\usepackage{stmaryrd}
\usepackage[colorlinks=true,breaklinks=true,linkcolor=red,anchorcolor=red,citecolor=red,urlcolor=red]{hyperref}
\usepackage{booktabs}
\usepackage{multirow}

\begin{document}

	\title{Dissipationless Layertronics in Axion Insulator $\mathbf{MnBi_2Te_4}$}
	\date{\today}
	
\author{Shuai Li}
\thanks{Shuai Li and Ming Gong are co-first authors}
\affiliation{School of Physical Science and Technology, Soochow University, Suzhou 215006, China.}
\affiliation{Institute for Advanced Study, Soochow University, Suzhou 215006, China.}

\author{Ming Gong}
\email{Corresponding author: minggong@pku.edu.cn}
\affiliation{International Center for Quantum Materials, School of Physics, Peking University, Beijing 100871, China}

\author{Shuguang Cheng}
\affiliation{Department of Physics, Northwest University, Xi’an 710069, China.}

\author{Hua Jiang}
\email{Corresponding author: jianghuaphy@suda.edu.cn}
\affiliation{School of Physical Science and Technology, Soochow University, Suzhou 215006, China.}
\affiliation{Institute for Advanced Study, Soochow University, Suzhou 215006, China.}
	
\author{X. C. Xie}
\affiliation{International Center for Quantum Materials, School of Physics, Peking University, Beijing 100871, China}
\affiliation{CAS Center for Excellence in Topological Quantum Computation, University of Chinese Academy of Sciences, Beijing 100190, China}

	\begin{abstract}
Surface electrons in axion insulators are endowed with a topological layer degree of freedom followed by exotic transport phenomena, e.g., the layer Hall effect [Gao \emph{et al.}, Nature
$\bm{595}$, 521 (2021)]. Here, we propose that such a layer degree of freedom can be manipulated in a dissipationless way based on the antiferromagnetic $\rm{MnBi_2Te_4}$ with tailored domain structure. This makes $\rm{MnBi_2Te_4}$ a versatile platform to exploit the ``layertronics'' to encode, process, and store information. Importantly, the layer filter, layer valve, and layer reverser devices can be achieved using the layer-locked chiral domain wall modes. The dissipationless nature of the domain wall modes makes the performance of the layertronic-devices superior to those in spintronics and valleytronics. Specifically, the layer reverser, a layer version of Datta-Das transistor, also fills up the blank in designing the valley reverser in valleytronics. Our work sheds light on constructing new generation electronic devices with high performance and low energy consumption in the framework of layertronics.     
	\end{abstract}

	\maketitle

\emph{Introduction}.--The invention and the applications of transistors have declared great success in manipulating the electronic charge degree of freedom \cite{RevModPhys.71.S336}. With the increasingly deepened understanding of electronic transport, concepts of designing high performance devices using internal degrees of freedom of electrons have sprung up. Among these, spintronics \cite{dotspinfil2000,baspinfil2001,spintrsis2003,spintrsis2019,spintrans2007,spintro2004,NobLec2008,tunnspingra2010,graspintron2014,supspin2015,spinval2011,spinval2019,topoantispin2018} and valleytronics \cite{vallyfil2007,vallcontgra2007,vallvalgra2008,vallhallgra2011,higwaygra2011,sponQHSgra2011,gapvalfil2011,magcontvall2013,pervallfil2015,vallfiteff2016,quanspinvall2021,valhallmos2016} are the best-known paradigms. Spintronic-devices such as the spin filter \cite{dotspinfil2000,baspinfil2001}, spin valve \cite{spinval2011,spinval2019}, and spin transistor \cite{spintrsis2003,spintrsis2019}, and valleytronic-devices such as the valley filter \cite{vallyfil2007,gapvalfil2011,pervallfil2015} and valley valve \cite{vallyfil2007,vallvalgra2008}, were theoretically raised. Tremendous experimental efforts have been devoted to realizing these promising devices in various materials \cite{valhallmos2016,IrMnspinfil2000,spfilvalgra2015,spintrsis2009,vallvalsplit2018}. Nevertheless, from the application level, the energy consumption strongly limits their performance. For many spintronic or valleytronic devices, the bulk carriers are inevitably scattered by the impurities, which dramatically lower their efficiency and increase their energy consumption \cite{vallfiteff2016,disspvalley1986,vallvalgra2008,vallspinpum2017}. It is thus highly desirable to exploit a new degree of freedom of electrons that is robust against disorder.

Recently, a new type of Hall effect, dubbed the layer Hall effect, was reported in even-layered antiferromagnetic (AFM) axion insulator (AI) $\rm{MnBi_2Te_4}$ \cite{LHE2021}. The layer-locked Berry curvature endows the surface electrons in the AI a topological nontrivial degree of freedom. Hopefully, one can construct devices using such a layer degree of freedom, and the encoded information can be easily read out through layer resovled transport measurements \cite{LHE2021,robAI2020}. However, the existed experimental and theoretical advances are unable to utilize topologically protected excitations to manipulate the layer degree of freedom in AFM $\rm{MnBi_2Te_4}$ dissipationlessly \cite{LHE2021,robAI2020,Magtran2022,layhidberry2022,intlaypol2022}. 

In this work, motivated by recent experimental progresses \cite{LHE2021,spinscaAI2019,topoAxsta2019,PreobsAntitopo2019,intrimagtopo2019,unithickAI2019,robAI2020,magimagAI2020,magcontAI2019,qahamagAI2020,gateahe2020,evenodd2021,magphatran2022,topspinmag2022,nonrectran2022,chiredgcond2017,lin_direct_2021,PhysRevB.105.165411,doi:10.1021/acsnano.2c03622,Toposurftemp2021,kataev2021}, we highlight that such a layer degree of freedom can be manipulated in a dissipationless way using the domain wall (DW) modes of the AFM $\rm{MnBi_2Te_4}$. In parallel to spintronics and valleytronics, we introduce the ``layertronics'', which is committed to designing scalable, low-dissipation devices to encode, process, and store information using the layer degree of freedom of AIs. Accordingly, we propose three of the most important layertronic-devices: layer filter, layer valve, and layer reverser.
Firstly, the layer filter can be constructed through a single DW of AFM $\rm{MnBi_2Te_4}$. 
We demonstrate that the injected layer-unpolarized current can be successfully filtered to be fully layer-polarized and transport dissipationlessly. The efficiency of the layer filter is robust against disorder and irregularity of the DW structures. Then, the layer valve can be achieved using two pairs of domains. By controlling the chemical potential of different domains, the layer valve can turn on or off the layer-polarized currents. Our results verify that the layer valve has high on-off ratio and is also robust against disorder. Finally, the layer reverser can be constructed utilizing the Chern insulator (CI) phase of the ferromagnetic (FM) $\rm{MnBi_2Te_4}$ \cite{robAI2020,magphatran2022}. The chiral DW mode in AI-CI-AI heterostructure connects the top and bottom surfaces, thus can reverse the layer-polarized current dissipationlessly. The layer reverser bears similarity to the Datta-Das transistor in spintronics \cite{datdas1990,dattadas12003,effmagdas2020}, and also fills up the blank of ``valley reverser'' in valleytronics.

\emph{Model Hamiltonian and the layer filter}.--We model the AFM $\rm{MnBi_2Te_4}$ by a 3D topological insulator (TI) with antiparallel layer magnetization \cite{topoAxsta2019,cribehaAI2021}. The Hamiltonian is $H=H_0+H_{\rm mag}$. $H_0$ represents a four-band TI  \cite{TImod12009,TImod22010} with
$H_0(k)=\sum^4_{i=1} d_{i}(\mathbf{k}) \Gamma_i,$
where $d_1(\mathbf{k})=A_1 k_x$, $d_2(\mathbf{k})=A_1k_y$, $d_3(\mathbf{k})=A_2k_z$ and $d_4(\mathbf{k})=M_0-B_1k^2_z-B_2(k^2_x+k^2_y)$. $\Gamma_i=s_i\otimes \sigma_1$ ($i=1,2,3$) and $\Gamma_4=s_0\otimes \sigma_3$ are the Dirac matrices. Hereafter, we choose $M_0=0.3$, $A_1=A_2=0.55$ and $B_1=B_2=0.25$ \cite{cribehaAI2021}. $H_{\rm mag}$ gives the Zeeman splitting 
$H_{\rm mag}=M(z) s_3\otimes \sigma_0,$
where $M(z)=\pm M_z$ is the magnetization of each individual layer along $z$ direction, which is fixed as $M_z=0.05$ if not other specified \footnote{In our calculations, the side surface states are gaped out in order to exclude their transport contribution for a realistic large-size $\rm{MnBi_2Te_4}$ sample}. For the AFM AI, the magnetization alignment of two adjacent layers are antiparallel. In the Supplementary Material \cite{supple}, we performed calculations using realistic material parameters with $M_0$=-0.12 eV, $A_1$=2.70 eV$\cdot {\rm{\AA}}$, $A_2$=3.20 eV$\cdot {\rm{\AA}}$, $B_1$=-11.90 eV$\cdot {\rm{\AA}^2}$, $B_2$=-9.40 eV$\cdot {\rm{\AA}^2}$, and $a$=1.35 nm \cite{topoAxsta2019,mobius2020,nonlocsurftr2021}. $M_z$=0.04 eV is adopted, which induces surface magnetic gap of about 0.08 eV and is in consistent with the theoretical and experimental studies (ranging from 0.05 to 0.10 eV) of $\rm{MnBi_2Te_4}$ \cite{spinscaAI2019,topoAxsta2019,PreobsAntitopo2019,Ji2021,Zeugner2019,Li2019,Macdonald2022}.
\begin{figure}
    \centering
    \includegraphics[width=85mm]{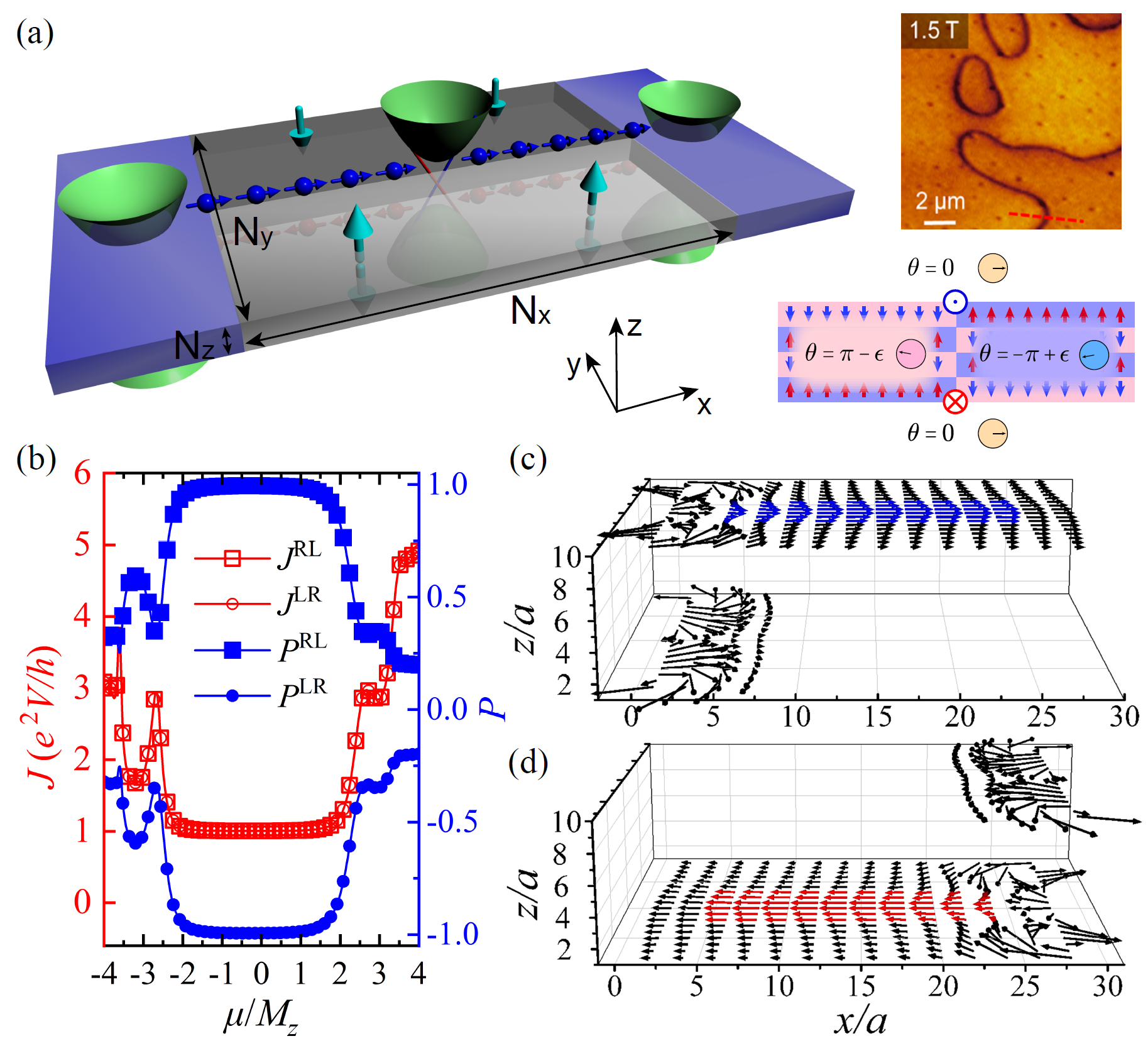}
    \caption{(color online) (a) Left panel: schematic of a layer filter. Green hyperbolic bands denote the dispersion of the gaped surface states. Red (blue) line represents the top (bottom) DW modes. Left and right terminals are AFM AIs with the Fermi energy $E_F$ located in the surface states. Right panel: the magnetic force microscopy image of the domain structure in AFM AI $\rm{MnBi_2Te_4}$ (adopted from Ref. \cite{magimagAI2020}), below which shows the sketch of the layer magnetization of AIs and the direction of chiral DW modes. The circled arrows on each domain indicate the axion angle $\theta$. (b) Transmission current $J$ and layer polarization $P$ of the layer filter vs chemical potential $\mu$ (in units of $M_z$) of the central region. (c) and (d) are local current distributions for top/bottom-layer filter ($\mu=0.02$), respectively. Blue (red) arrows emphasize conducting current near the DWs. The model parameters are  $N_z=10$, $N_y=N_x=20$, chemical potential of the terminals is $0.2$, and $a$ is the lattice constant which measures the distance between two adjacent septuple layers.}
    \label{fig:filter1}
\end{figure}

The layer filter is the most fundamental element of layertronic devices, which generates fully layer-polarized current. As sketched in Fig. \ref{fig:filter1}(a), the layer filter is constructed through a single DW of the AFM $\rm{MnBi_2Te_4}$. The magnetization direction is illustrated by the cyan arrows. Due to the AFM magnetization, gaps are opened on Dirac surface states of the AI, leading to half-quantized Hall conductances $\pm e^2/2h$ on the top or bottom surfaces \cite{halfquan2011,PhysRevB.98.245117,supple}. The AFM order introduces topological magneto-electric coupling term $\Delta \mathcal{L}=\frac{\alpha}{4\pi^2} \theta\mathbf{E}\cdot\mathbf{B}$, where $\alpha$ is the fine structure constant, and $\theta$ the axion angle \cite{topofiel2008,magelecpol2009,axielecdya2021}. When the two different types of AFM orders meet to form a heterostructure, the metallic interface states of the AIs are gapped \cite{zhou_transport_2022,gong_half-quantized_2022}, leaving the bulk insulated. The crossing lines of the three phases, i.e., the two types of AFM AIs (with $\theta=\pm \pi$) and the vacuum ($\theta=0$), can be viewed as line defects of the $\theta$ field \cite{RevModPhys.88.035005}. Its winding direction is illustrated in Fig. \ref{fig:filter1}(a). Consequently, chiral modes appear on the vortices of $\theta$ due to the Callan-Harvey anomaly \cite{callan_anomalies_1985}. These chiral DW states are locked with the layer degree of freedom and facilitate the generation of layer-polarized transmission current.  

The AFM DWs can be easily achieved experimentally in $\rm{MnBi_2Te_4}$, as shown by the magnetic force microscopy image of the domain structure in AFM AI $\rm{MnBi_2Te_4}$ [right panel of Fig. \ref{fig:filter1}(a)]\cite{magimagAI2020} \footnote{In bilayer graphene,  similar valley filter devices have been realized in naturally formed DWs \cite{Ju2015}}. Besides, the AFM order in $\rm{MnBi_2Te_4}$ can be reversed through the $\mathbf{E}\cdot\mathbf{B}$ term \cite{LHE2021}, meaning that it is possible to fabricate the DWs in a gate-controllable manner \cite{supple}. The external lead is realized by locating Fermi energy $E_F$ of the AI into the surface bands, but kept inside the bulk gap to ensure that only surface currents are engaged. To quantitatively investigate the layer polarization of the current, we define the polarization coefficient as \cite{vallyfil2007,vallfiteff2016} 
\begin{equation}
P(E_F)=\frac{J_t(E_F)-J_b(E_F)}{J_t(E_F)+J_b(E_F)},
\end{equation}
where the layer-resolved currents are given by
\begin{equation}
J_{t(b)}(i_x,E_F)=\sum_{\begin{tiny}{\begin{matrix}&0\leqslant i_y\leqslant N_y,\\ &i_z>(\leqslant)N_z/2\end{matrix}}\end{tiny}} J_{x}(i_x,i_y,i_z,E_F).
\end{equation}
Here, the layer filter considered is a slab with size $N_x\times N_y\times N_z$ [Fig. \ref{fig:filter1}(a)]. $J_{x}(i_x,i_y,i_z,E_F)$ is the $x$-component of the local current on site $(i_x,i_y,i_z)$ under the bias $V$, and $J_t$($J_b$) represents the transmission current near the top (bottom) layers. We numerically calculate the transmission current $J=J_t+J_b$ and layer polarization $P$ versus chemical potential $\mu$ of the central region, where $\mu=E_F-E_0$ measures the difference between $E_F$ and the energy of the Dirac point $E_0$ [red line with empty squares and blue line with filled squares in Fig. \ref{fig:filter1}(b)]. In the gap of the surface states, $J=e^2V/h$ is quantized, meaning that the filting process is dissipationless and the transmission current becomes fully layer-polarized on the drain terminal with $P=+1$. We further calculate the local current distribution to visualize the layer polarization. As shown in Fig. \ref{fig:filter1}(c), the layer filter only permits the top-layer current (indicated by the blue arrows) while blocks the bottom layer current, and we call it the top-layer filter (TLF). Similarly, we can also obtain the bottom-layer filter (BLF) simply by reversing the source and drain. $J$ and $P$ [Fig. \ref{fig:filter1}(b)] show that inside the gap, the transmission current is still quantized ($J=e^2V/h$), but fully layer-polarized with $P=-1$. The local current distribution also verifies that the current only goes through the bottom surface, as shown in Fig. \ref{fig:filter1}(d). The layer polarization of the current can be detected through the scanning microwave impedance microscopy images. The measured current signal on the top layer of $\rm{MnBi_2Te_4}$ for the TLF will be much stronger than that of the BLF, although the conductances for the two cases are both quantized \cite{lin_direct_2021,PhysRevB.105.165411,Yang2022}. 

\begin{figure}
\centering
\includegraphics[width=1\linewidth]{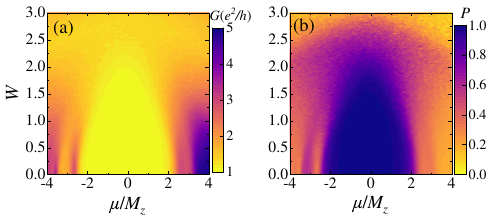}
\caption{(color online) (a) and (b) are $G$ and $P$ of the TLF as functions of $\mu$ and $W$, respectively. Other  parameters are the same as that in Fig. \ref{fig:filter1}}.
\label{fig:filter2}
\end{figure}

The performance of electronic devices is usually suppressed by disorders \cite{vallfiteff2016,vallspinpum2017}. Here, we show that the efficiency of the layer filter is robust against weak disorders. We introduce a random onsite potential term 
$H_{\rm dis}=U(r)~s_0\otimes \sigma_0$, where $U(r)$ is a local potential uniformly distributed within $[-W/2,W/2]$. The differential conductance $G=J/V$ and layer-polarization $P$ of the TLF versus $\mu$ and disorder strength $W$ are plotted in Fig. \ref{fig:filter2}. For small $W$, $G$ remains quantized and the current maintains fully polarized ($P=1$) for a $\mu$-window. By increasing $W$, the $\mu$-windo with quantized $G$ for the TLF shrinks, but still survives under strong disorder strength (such as $W=1$, see Fig. \ref{fig:filter2}) due to the robustness of the topologically protected chiral DW modes.
	
Though the key physics are captured by the model study, more realistic considerations can facilitate the experimental realization of the layer filter \cite{supple}. Calculations using realistic material parameters of $\rm{MnBi_2Te_4}$ show that an ideal DW (the DW width $W_D$=0) can act as a filter when $E_F$ is inside the surface gap ($\approx2M_z$=0.08 eV). The finite width of the DW can induce trivial one-dimensional sub-bands, of which the gap is proportional to $1/W_D$ and may lie within the surface gap. Nevertheless, as we estimated using realistic material parameters, a sub-band gap larger than 10 meV can be observed for $W_D\approx$ 500 nm, which is able to be distinguished by current transport experiments. Moreover, the non-uniformity, the different spin-flop structure, as well as the curved shape of the DWs have little influence on the quantized transport of the layer filter \cite{supple}. Therefore, the dissipationless transport $G=e^2/h$ and the full layer-polarization $P=\pm1$, are robust against the irregularity of the DWs in $\rm{MnBi_2Te_4}$.

\begin{figure}
    \centering
    \includegraphics[width=85mm]{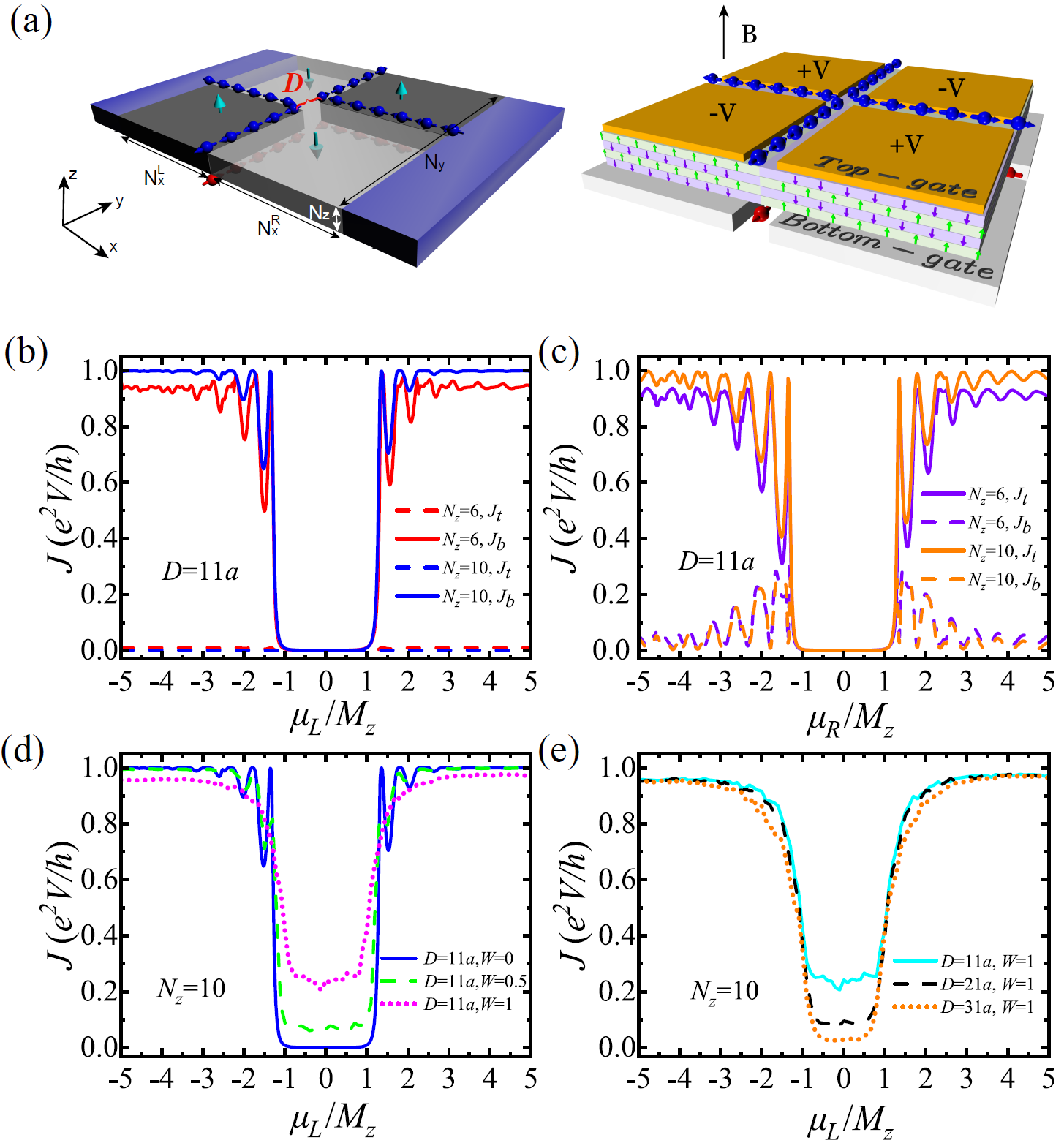}
    \caption{(color online) (a) Left panel: schematic of the layer valve. The DWs are separated by $D$ in units of the lattice constant $a$. Right panel: the domain structure required by the layer valve can be realized in a gate-controlled way under a perpendicular magnetic field $\mathbf{B}$\cite{LHE2021}. (b)-(c) Layer-resolved currents $J_{t}$ and $J_{b}$ of the valve vs the TLF (BLF) chemical potential $\mu_{L}$ ($\mu_{R}$), with fixed $\mu_{R}=0.02$ ($\mu_{L}=0.02$) of BLF (TLF), respectively. (d)-(e) $J=J_{t}+J_{b}$ in the presence of disorders vs $\mu_{L}$ ($\mu_{R}=0.02$) for different (d) $W$ and (e) $D$. Model parameters: $N_y=50$, $N^{L}_{x}=N^{R}_{x}=50$ and terminal chemical potential is $0.25$.}
    \label{fig:valve1}
\end{figure}

\emph{Layer valve}.--The layer valve device serves as a switch to turn ``on'' or ``off'' the specific layer-polarized current. It is designed by connecting the TLF and BLF sequentially from left to right as sketched in Fig. \ref{fig:valve1}(a). 
 $\mu_{L}$ ($\mu_{R}$) is the chemical potential of the TLF (BLF) in the left (right) part of the valve. Figures \ref{fig:valve1}(b) and (c) show that the layer-resolved currents $J_t$ and $J_b$ [versus $\mu_{L}$ ($\mu_{R}$) of the TLF (BLF)] both vanish in the gap of the surface states, indicating zero transmission current. 
The layer valve is in the ``off'' status. This can be ascribed to the opposite chirality of the DW modes in TLF and BLF [Fig. \ref{fig:valve1}(a)]. The chirality reversal turns off the valve by blocking the transmission current on both the top and bottom layers. When we independently shift $\mu_{L}$ ($\mu_{R}$) to the surface states, the layer valve is turned ``on" and generates current with bottom (top) layer polarization. As shown in Fig. \ref{fig:valve1}(b), by shifting $\mu_{L}$, $J_{b}$ grows up while $J_{t}$ keeps zero, indicating that the valve is turned on and works as a BLF. Similarly, when $\mu_{R}$ is tuned such that $E_F$ locates deeply inside the surface states, $J_{b}$ keeps small while $J_{t}$ becomes significantly large [Fig. \ref{fig:valve1}(c)], which implies that the valve is turned on as a TLF. 
Moreover, the quantized transport of $J_{b}$ and $J_{t}$ are better for thicker valves [see curves in Figs. \ref{fig:valve1}(b) and (c) with $N_z=6$ and $N_z=10$], which originate from the suppression of the backscattering between spatially seperated DW modes.

The performance of the layer valve is also robust against disorders. As shown in Fig. \ref{fig:valve1}(d), when $\mu_{L}$ is tuned such that $E_F$ lies inside the gap of the TLF and $W<0.5$, $J$ slightly deviates from zero, indicating that the layer valve is robust against weak disorders. For strong disorders, e.g., $W=1$, $J$ climbs up, leading to the decreaing of the on-off ratio. This stems from the scattering of DW modes in TLF to that in BLF under disorders. However, the on-off ratio can be significantly enhanced by separating the DW in TLF and BLF further [increasing $D$, see Fig. \ref{fig:valve1}(a)], which suppresses the scattering. In Fig. \ref{fig:valve1}(e), $J$ is lowered as $D$ increases, thus the on-off ratio rises accordingly. For spintronics/valleytronics, disorder induces large scattering between opposite spin/valley components, which greatly suppresses the on-off ratio of the spin/valley valve \cite{vallvalgra2008,spinrelax1999}. In layertronics, besides the topologically protected chiral DW modes, the high performance of layer valve under disorder is also ensured by the spatially resolved layer degree of freedom.    
These transport properties are also verified by using realistic material parameters of $\rm{MnBi_2Te_4}$ \cite{supple}.

Experimentally, the layer valve can be achieved in a gate-controlled way. As shown in Fig. \ref{fig:valve1}(a), under a perpendicular magnetic field $\mathbf{B}$, the applied gate voltages on the four split-gates induce alternative electric field $\mathbf{E}$, giving rise to the domain structure of the AFM order due to $\Delta \mathcal{L}=\frac{\alpha}{4\pi^2} \theta\mathbf{E}\cdot\mathbf{B}$ of the AI. Therefore, the layer valve devices with high on-off ratio are experimentally feasible.

\begin{figure}
	\centering
	\includegraphics[width=85mm]{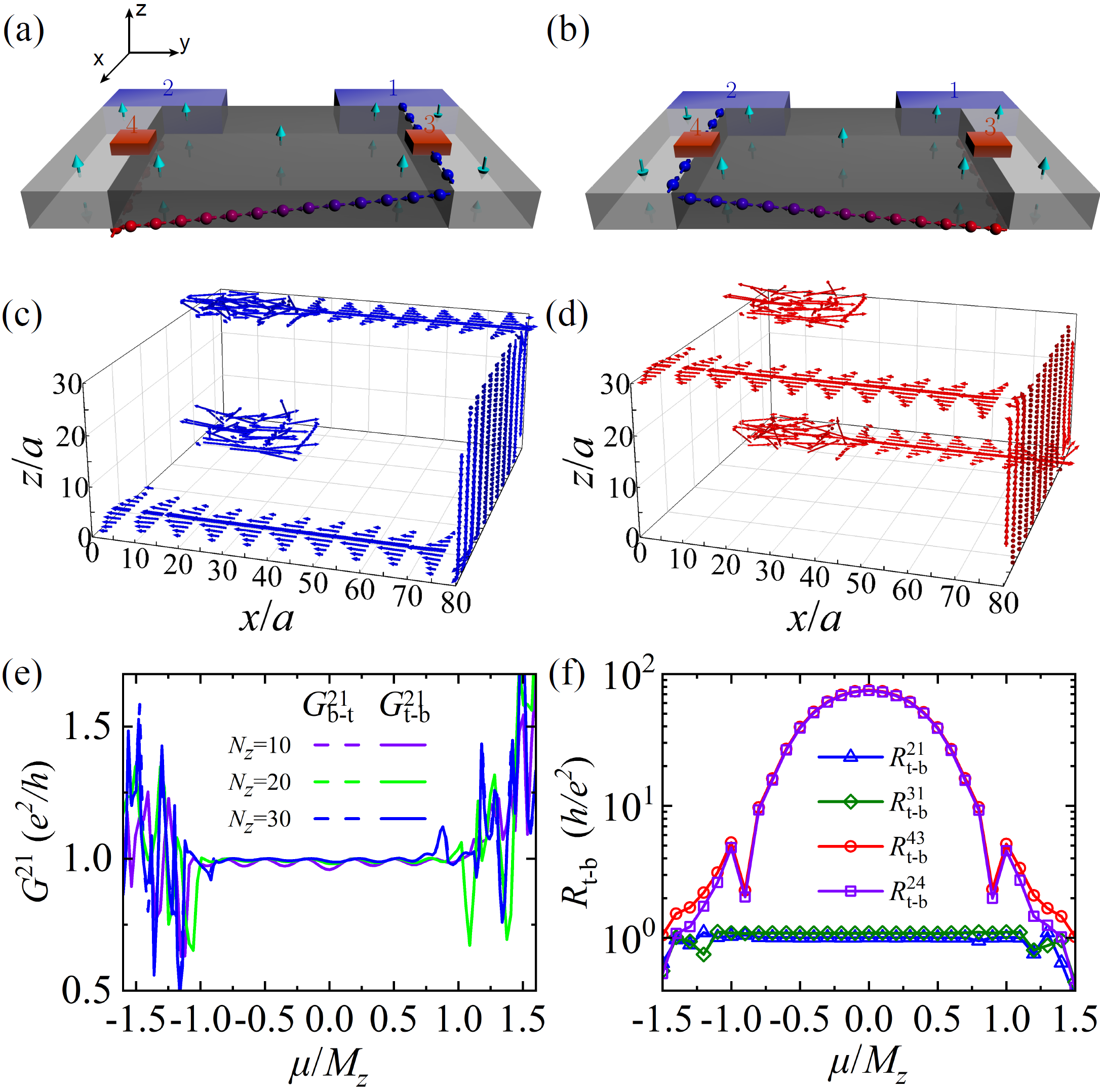}
	\caption{(color online) Schematics of top-bottom (a) and bottom-top (b) layer reversers, respectively. Colored arrows mark the layer reversing of the transmission modes. Terminals 3 and 4 are metallic contacts on the top surface. (c)-(d) Local current distributions of the two kinds of reversers ($N_z=30$), with $\mu=0.02$. (e) Differential conductance between terminals 1 and 2 of top-bottom reverser $G^{21}_{\rm t-b}$ and bottom-top reverser $G^{21}_{\rm b-t}$ vs $\mu$ with terminals 3 and 4 floated. (f) Local resistances of the top-bottom reverser vs $\mu$. Model parameters: $N_y=108$, $N_x=60$, terminal chemical potential is 0.2 and the size of contacts 3 and 4 is $8\times 8$.}
	\label{fig:reverser1}
\end{figure}

\emph{Layer reverser}.--In applications, one expects that the binary bits stored inside the logical unit could be switched easily. However, switching internal degrees of freedom of electrons is challenging in experiments, especially for spintronics and valleytronics. An example is that the Datta-Das transistor \cite{datdas1990,dattadas12003,effmagdas2020}, the reverser of the spin degree of freedom of electrons, has not been completely realized and applied since its first proposal. Similar dilemma appears in valleytronics where efficient valley reverser is hard to design. Interestingly, the feasibility of tuning the AFM AI $\rm{MnBi_2Te_4}$ into the ferromagnetic (FM) CI phase facilitates the layer reverser device, which inverts the layer degree of freedom dissipationlessly. As sketched in Fig. \ref{fig:reverser1}(a) and (b), the reverser consists of a FM CI domain sandwiched by two AFM AIs with different AFM orders, which can be achieved experimentally by applying vertical magnetic field on $\rm{MnBi_2Te_4}$ to 
reform the magnetization alignment \cite{robAI2020,magphatran2022}. 

The key ingredient that triggers the layer reversing is the layer-crossing chiral edge mode, which starts from the top/bottom surface, goes through the CI edge mode on the side surface, then flows back into the bottom/top surface. The local current distributions [Fig. \ref{fig:reverser1}(c)] show the top layer current is inverted to the bottom layer. We denote such a reverser as a top-bottom layer reverser. Similarly, the bottom-top reverser is realized by simply reversing the AFM order of the AIs \cite{LHE2021} [Fig. \ref{fig:reverser1}(b)(d)]. The dissipationless nature of the layer reverser is reflected by the quantization of the conductances between terminals 1 and 2 $G^{21}_{\rm t-b}$ and $G^{21}_{\rm b-t}$ for the two kinds of reversers. As shown in Fig. \ref{fig:reverser1}(e), in the gap of the surface states, $G^{21}_{\rm t-b}$ and $G^{21}_{\rm b-t}$ are nearly quantized ($e^2/h$). As the thickness $N_z$ grows from 10 to 30, $G^{21}_{\rm t-b}$ and $G^{21}_{\rm b-t}$ are better quantized due to the suppression of the backscattering between two DW modes.

Experimentally, the reversal of layer polarization, as well as the information encoded by the layer degree of freedom, can be detected through the layer-resolved transport measurements. It can be achieved by attaching two additional point contacts (terminal 3 and 4) at the DWs on the top surface [Fig. \ref{fig:reverser1}(a)(b)], and measuring the resistances between different pairs of terminals. As shown in Fig. \ref{fig:reverser1}(f), for the top-bottom reverser, colossal local resistances $R^{43}_{\rm t-b}$ and $R^{24}_{\rm t-b}$ appear around $\mu=0$ because of the absence of transmission mode that connects terminal 4 with the others. In contrast, resistances $R^{21}_{\rm t-b}$ and $R^{31}_{\rm t-b}$ are quantized $h/e^2$ because of the chiral mode that connects terminals 1 and 2 or 1 and 3. 

\emph{Conclusion}.--We introduced the layertronics and designed key devices utilizing the layer degree of freedom of the AFM AI $\rm{MnBi_2Te_4}$. The corresponding layer filter, layer valve devices were constructed based on the AFM DW modes. The layer reverser device, which circumvents the difficulties in constructing reversers in spintronics and valleytronics, can be achieved through the FM CI phase of $\rm{MnBi_2Te_4}$. The requirements for realizing layertronics are achievable under state-of-the-art experimental conditions. The robustness of DW modes under disorders realizes dissipationless manipulation of the layer degree of freedom, making layertronics a promising paradigm in constructing new generation electronic devices.

\begin{acknowledgments}
We thank Weida Wu, Jian Shen, Fengqi Song, Yijia Wu, and Zhiqiang Zhang for fruitful discussions. This work is financially supported by the National Basic Research Program of China (Grants No. 2019YFA0308403 and No. 2015CB921102), the National Natural Science Foundation of China (Grants No. 11822407 and No. 11874298), and the Strategic Priority Research Program of the Chinese Academy of Sciences (Grant No. XDB28000000).
\end{acknowledgments}

\bibliography{ref}

\end{document}


\title{Supplementary Materials for ``Dissipationless Layertronics in Axion Insulator $\bf{MnBi_2Te_4}$"}

\author{Shuai Li}
\thanks{Shuai Li and Ming Gong are co-first authors}
\affiliation{School of Physical Science and Technology, Soochow University, Suzhou 215006, China.}
\affiliation{Institute for Advanced Study, Soochow University, Suzhou 215006, China.}

\author{Ming Gong}
\email{Corresponding author: minggong@pku.edu.cn}
\affiliation{International Center for Quantum Materials, School of Physics, Peking University, Beijing 100871, China}

\author{Shuguang Cheng}
\affiliation{Department of Physics, Northwest University, Xi’an 710069, China.}

\author{Hua Jiang}
\email{Corresponding author: jianghuaphy@suda.edu.cn}
\affiliation{School of Physical Science and Technology, Soochow University, Suzhou 215006, China.}
\affiliation{Institute for Advanced Study, Soochow University, Suzhou 215006, China.}
	
\author{X. C. Xie}
\affiliation{International Center for Quantum Materials, School of Physics, Peking University, Beijing 100871, China}
\affiliation{CAS Center for Excellence in Topological Quantum Computation, University of Chinese Academy of Sciences, Beijing 100190, China}

\maketitle
\renewcommand{\theequation}{S\arabic{equation}}
\renewcommand{\thefigure}{S\arabic{figure}}
\renewcommand{\thesection}{S\Roman{section}}

\tableofcontents

\section{Physical origin of the chiral domain wall modes in antiferromagnetic axion insulator $\mathbf{MnBi_2Te_4}$}

\begin{figure}
\centering
\includegraphics[width=14cm]{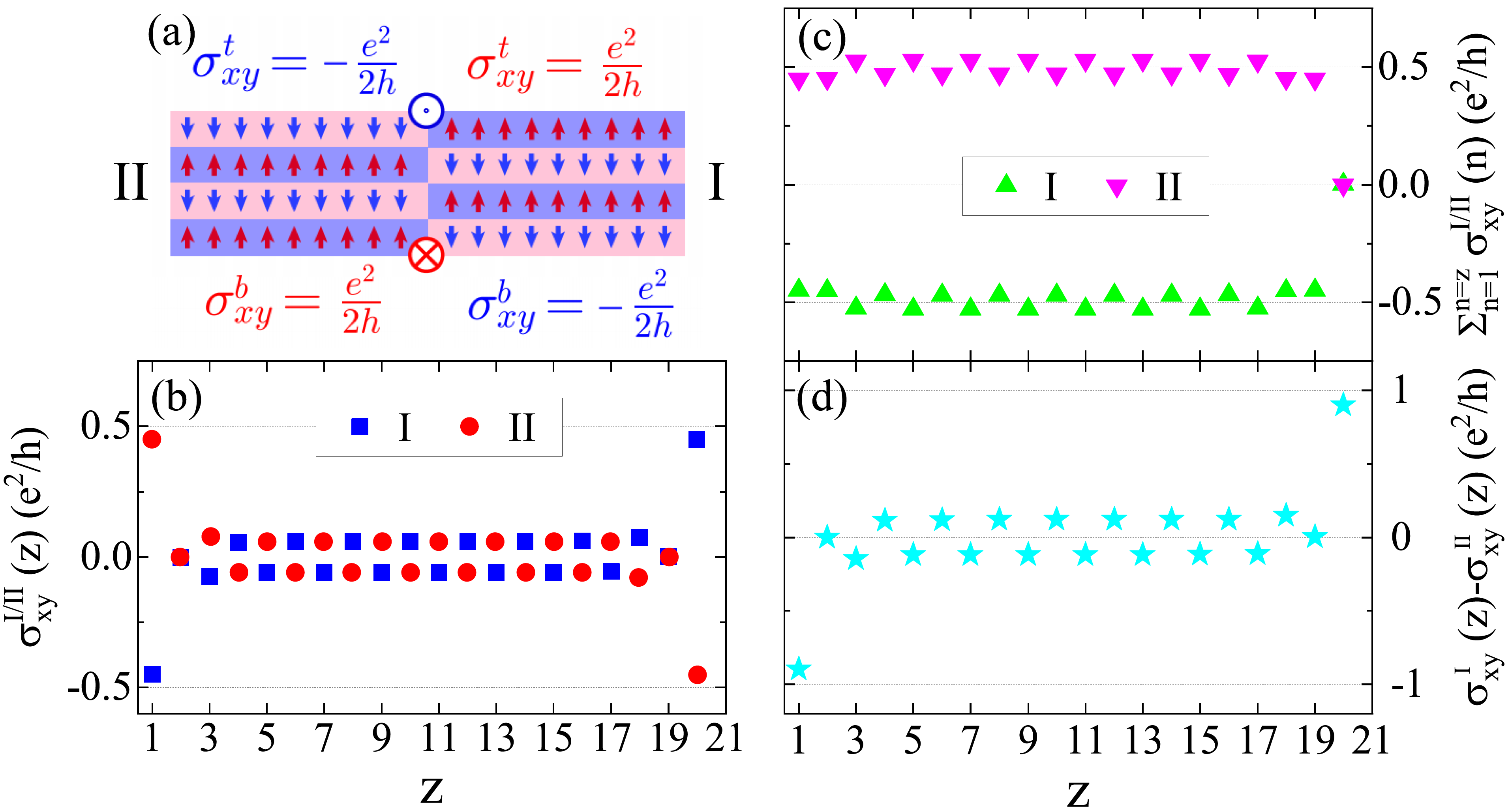}
\caption{(color online) (a) Sketch of the magnetic DW between the two AFM ground states I and II of $\mathrm{MnBi_2Te_4}$. Blue and red arrows illustrate the  magnetization in each single septuple layers. The vertices at the interface represent the propagation direction of the chiral DW modes. The half-quantized surface Hall conductances are labeled correspondingly. (b) Layer-dependent Hall conductances $\sigma^{\rm{I/II}}_{xy} (z)$ of AFM states I and II, versus the layer index $z$. (c) The cumulative summation of the layer-dependent Hall conductances $\sigma^{\rm{I/II}}_{xy}(z)$ from the $1$st to the $z$th layer versus $z$. (d) Hall conductance difference between the two AFM states I and II as a function of $z$. Model parameters are: $M_z=0.05$, $N_z=20$ and Fermi energy of the AFM $\mathrm{MnBi_2Te_4}$ $E_F=0$.}
\label{sfig1}
\end{figure}

In this section, we show the physical origin of the chiral domain wall (DW) modes in antiferromagnetic (AFM) topological axion insulator (AI) $\mathrm{MnBi_2Te_4}$ in the view of Hall conductance. The two AFM ground states (I and II) of $\mathrm{MnBi_2Te_4}$ are sketched in Fig. \ref{sfig1}(a), where a magnetic DW is formed at the interface between them. We calculate the layer-dependent Hall conductance for AFM states I and II as functions of layer indices $z$ [see Fig. \ref{sfig1}(b)]. The Hall conductances on the inner layers ($z=2,\cdots,19$) are quantitatively small and oscillate around zero. While on the surface layers $z=1$ and $z=20$, the Hall conductances nearly approach $0.5 (e^2/h)$ but with a sign reverse as labeled in Fig. \ref{sfig1}(a), indicating that AFM $\mathrm{MnBi_2Te_4}$ owns a half-quantized layer Hall effect where electrons on the top and bottom surface layers deflect in opposite directions \cite{LHE2021}. We also evaluate the cumulative summation of the Hall conductance from each layer, i.e., $\sum^{n=z}_{n=1} \sigma^{\rm{I/II}}_{xy}(n)$, as given in Fig. \ref{sfig1}(c). The cumulative conductances maintain almost half-quantized from $z=1$ to $z=19$ and vanish at $z=20$, since the top and bottom surface layers cancel out each other. Therefore, the AFM AI $\mathrm{MnBi_2Te_4}$ is characterized by a zero Hall conductance plateau \cite{AICIphase2020}.
We give the difference of the layer Hall conductances between the two AFM states, $\sigma^{\rm{I}}_{xy}(z)-\sigma^{\rm{II}}_{xy}(z)$, as a function of layer indices $z$ in Fig. \ref{sfig1}(d). On the surfaces ($z=1,20$), the conductance differences are quantized ($\pm e^2/h$), giving rise to the generation of chiral modes at the DW between the two AFM domains. Inside the bulk ($z=2,\cdots, 19$), the Hall conductance differences are nearly zero. It indicates that the chiral DW modes are spatially separated on the two surfaces. And the chirality of the DW modes is determined by the sign ($\pm$) of the Hall conductance differences on the surfaces and thus locked to the layer degree of freedom. Therefore, the DW modes have 1D features as illustrated in Fig. \ref{sfig1}(a). Our numerical results give another view of the physical origin of the topological DW modes counter-propagating on different surfaces along the DW between the AFM $\mathrm{MnBi_2Te_4}$ domains I and II.

\section{Simulation of the layertronic devices with the realistic material parameters of $\bf{MnBi_2Te_4}$}
The numerical calculations in the main text for the ``layertronics" devices are based on a theoretical model with scaled parameters $A_1$, $A_2$, $B_1$, $B_2$, $M_0$ and perpendicular magnetization $M_z$ (in unit of $\frac{\hbar V_F}{a}$ where $V_F$ is the Fermi velocity of the Dirac surface states and $a=1.35$ nm which measures the distance between two adjacent septuple layers of $\rm{MnBi_2Te_4}$). 
To identify that our proposals are experimentally feasible in AFM AI $\rm{MnBi_2Te_4}$, we give in this section the complementary calculations and discussions that utilize the realistic material parameters of $\rm{MnBi_2Te_4}$.

\begin{figure}[h]
	\centering
	\includegraphics[width=11cm]{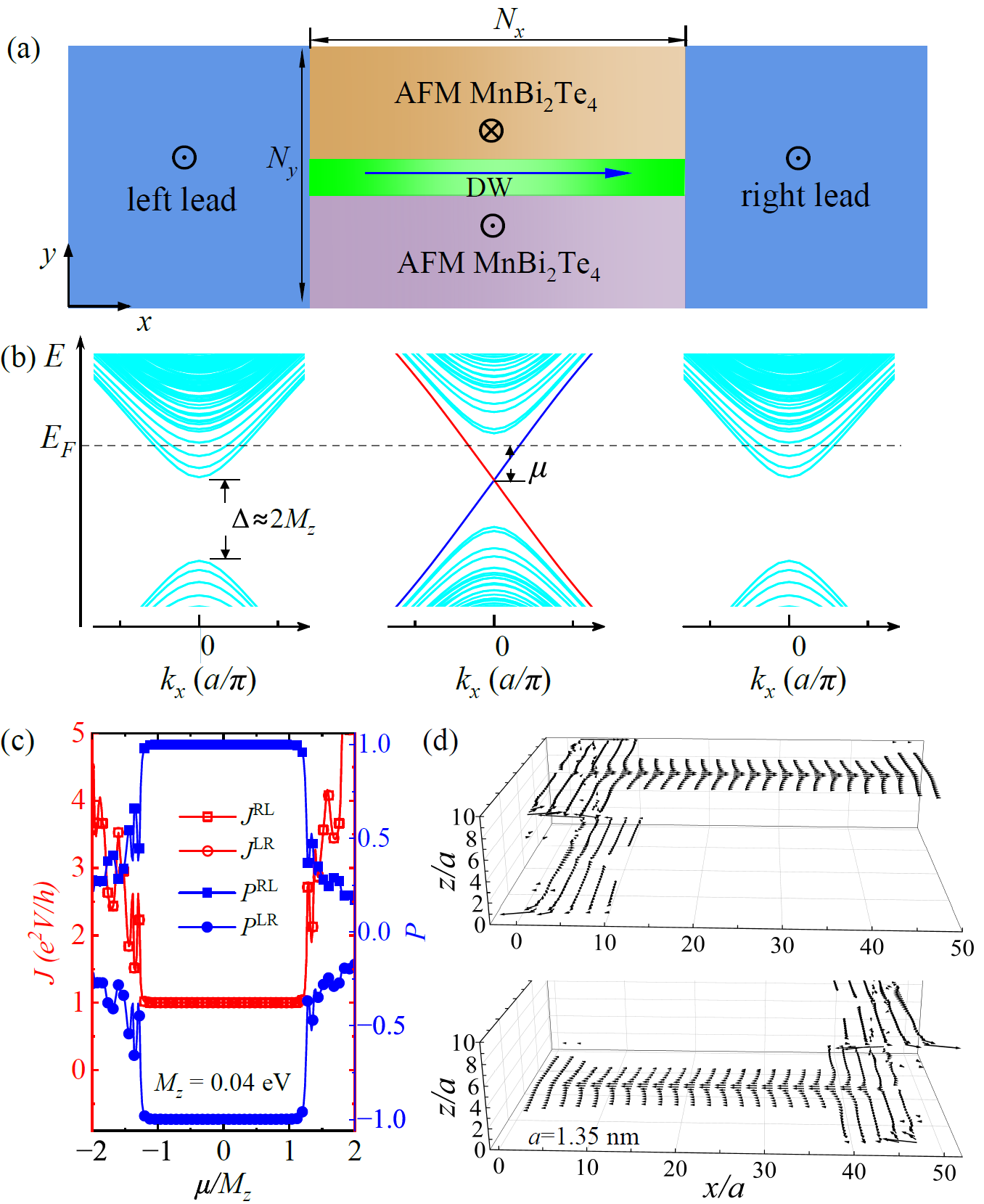}
	\caption{(color online) (a) Top-view of the two-terminal layer filter device constructed with AFM $\rm{MnBi_2Te_4}$. The vertices illustrate the directions (in and out) of the surface magnetization $M_z$. Horizontal blue arrow denotes the chiral DW mode on the top surface. (b) Corresponding energy spectrum of the leads and the filter. The blue/red linear dispersion highlights the DW mode on the top/bottom surface. $\mu$ is the chemical potential of the filter which measures the difference between Fermi energy $E_F$ and the Dirac point. (c) Transmission current $J^{RL}$ ($J^{LR}$) through the filter and its layer-polarization $P^{RL}$ ($P^{LR}$) versus $\mu/M_z$. The left/right (right/left) lead is the source/drain terminal. (d) Local current distributions of the top-layer (top panel) and bottom-layer (bottom panel) polarized currents through the layer filter, with $\mu/M_z=0.25$. Other model parameters are: $M_0$=-0.1165 eV, $A_1$=2.7023 eV$\cdot {\rm{\AA}}$, $A_2$=3.1964 eV$\cdot {\rm{\AA}}$, $B_1$=-11.9048 eV$\cdot {\rm{\AA}^2}$, $B_2$=-9.4048 eV$\cdot {\rm{\AA}^2}$, $M_z$=0.04 eV, $N_z$=10, $N_y$=40, $N_x$=40, lead chemical potential 2$M_z$=0.08 eV, and lattice constant $a$=1.35 nm ($a$ measures the distance between two adjacent septuple layers of $\rm{MnBi_2Te_4}$).}
	\label{sfig8}
\end{figure}

An effective model for AFM $\rm{MnBi_2Te_4}$ \cite{nonlocsurftr2021} is employed whose parameters are determined by fitting the energy spectrum of the effective Hamiltonian with that of the $ab$ initio calculation. The effective material parameters are given as $M_0$=-0.1165 eV, $A_1$=2.7023 eV$\cdot {\rm{\AA}}$, $A_2$=3.1964 eV$\cdot {\rm{\AA}}$, $B_1$=-11.9048 eV$\cdot {\rm{\AA}^2}$, $B_2$=-9.4048 eV$\cdot {\rm{\AA}^2}$ \cite{nonlocsurftr2021,TAImag2019,mobius2020}. Note that this effective model provides a reasonable perpendicular magnetization about $M_z$=0.04 eV along $z$ direction in $\rm{MnBi_2Te_4}$ septuple layers. Therefore, the surface band gaps of $\rm{MnBi_2Te_4}$ induced by intrinsic magnetism can be large of about $2M_z$=0.08 eV, as consistent with the $ab$ initio calculation \cite{TAImag2019}. The effective model is regularized on a cubic lattice with the lattice constant $a=1.35$ nm standing for the distance between two adjacent septuple layers of $\mathrm{MnBi_2Te_4}$. 

\begin{figure}[h]
	\centering
	\includegraphics[width=13cm]{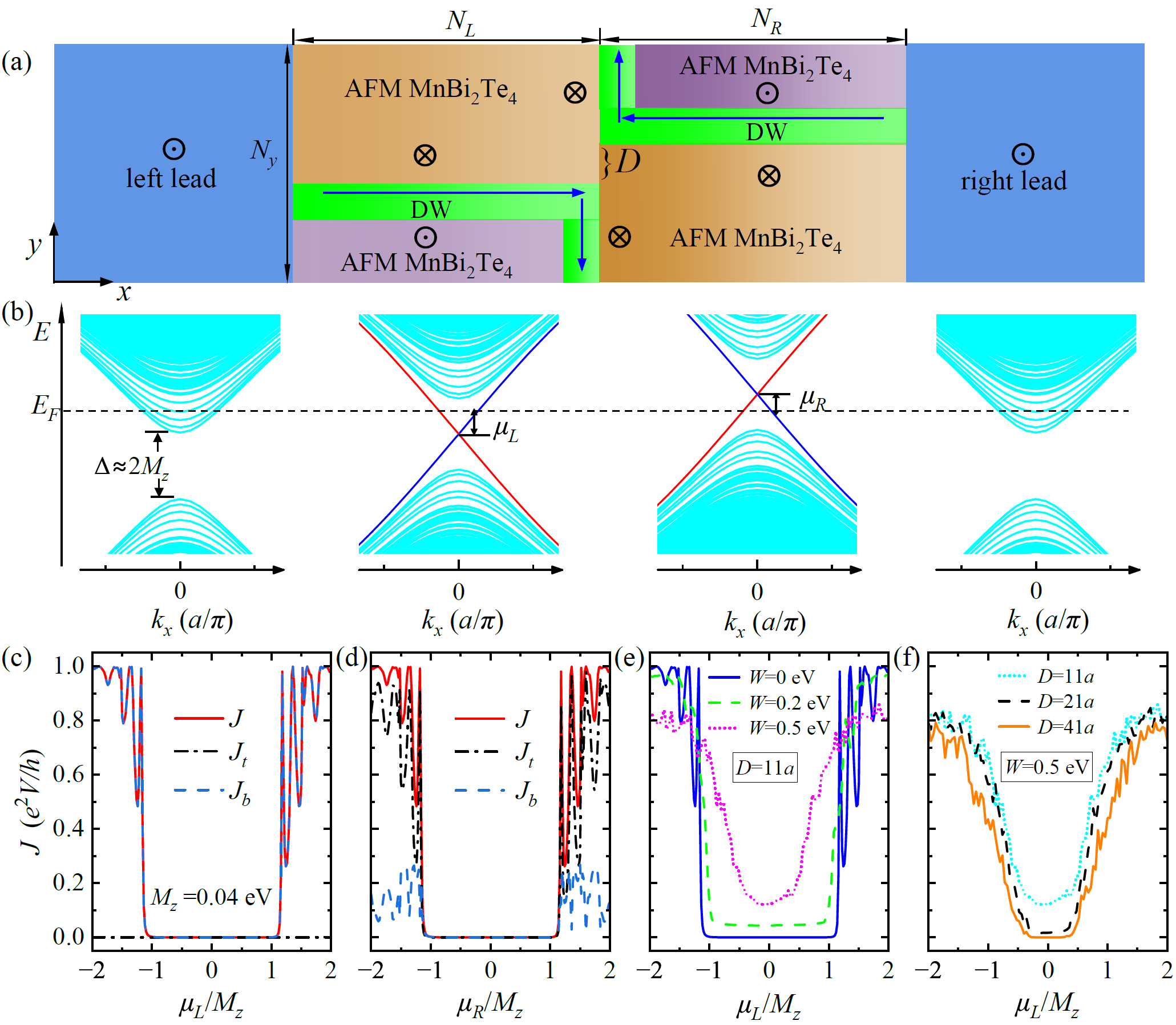}
	\caption{(color online) (a) Top-view of the two-terminal layer valve device fabricated with AFM $\rm{MnBi_2Te_4}$. The vertices represent the directions (in and out) of the surface magnetization $M_z$. Horizontal blue arrows denote the chiral DW modes on the top surfaces of the TLF and BLF. $D$ is the separation of the two DW modes in $y$ direction. (b) Corresponding energy spectrum of the leads and the valve. $\mu_{L}$ ($\mu_{R}$) is chemical potential of the TLF (BLF) that measures the difference between Fermi energy $E_F$ and the Dirac point of the DW modes in the TLF (BLF). The layer-resolved transmission currents $J_{t}$, $J_{b}$ and $J=J_{t}+J_{b}$ through the valve versus (c) $\mu_{L}/M_z$ with $\mu_{R}/M_z=0.25$ and versus (d) $\mu_{R}/M_z$ with $\mu_{L}/M_z=0.25$. (e) Transmission current $J$ through the valve versus $\mu_{L}/M_z$ with ($W=0.2$ eV and 0.5 eV) and without ($W$=0 eV) disorders. $\mu_{R}/M_z=0.25$. (f) Current $J$ through the valve versus $\mu_{L}/M_z$ for $D=11a$, $21a$ and $41a$, under disorders ($W=0.5$ eV). $\mu_{R}/M_z=0.25$. The left/right lead is the source/drain terminal.  Other model parameters are: $M_0$=-0.1165 eV, $A_1$=2.7023 eV$\cdot {\rm{\AA}}$, $A_2$=3.1964 eV$\cdot {\rm{\AA}}$, $B_1$=-11.9048 eV$\cdot {\rm{\AA}^2}$, $B_2$=-9.4048 eV$\cdot {\rm{\AA}^2}$, $M_z$=0.04 eV, $N_z$=10, $N_y$=50, $N_L$=50, $N_R$=50, lead chemical potential 2$M_z$=0.08 eV, and lattice constant $a$=1.35 nm.}
	\label{sfig9}
\end{figure}

The calculation results for the layer filter using realistic parameters are given in Figs. \ref{sfig8}. The energy spectrum of the device exhibits a sizable surface gap of $\Delta\approx 2M_z$ =0.08 eV [see Fig. \ref{sfig8} (b)]. Inside the surface gap of the layer filter, two linear and layer-locked DW modes emerge. When the Fermi level $E_F$ crosses over the surface states of the two leads and the DW states [see dashed line in Fig. \ref{sfig8}(b)], a layer-polarized and quantized transmission current can be induced. Figure \ref{sfig8}(c) plots the transmission current $J$ and its layer polarization $P$ versus the chemical potential $\mu$ of the filter. When $E_F$ locates inside the surface gap or at the DW states ($-1< \mu/M_z < 1$), a quantized current $J^{RL}=e^2V/h$ is flowing through with top-layer polarization $P^{RL}=+1$ [see the solid lines with empty and filled squares in Fig. \ref{sfig8}(c)]. The filtered current can be directly visualized by the local current distribution, as shown in the top panel of Fig. \ref{sfig8}(d). On the contrary, a bottom-layer polarized ($P^{LR}=-1$) current $J^{LR}=e^2V/h$ is transmitted through the filter by simply reversing the source and drain leads [see the solid lines with empty and filled circles in Fig. \ref{sfig8}(c) and bottom panel of Fig. \ref{sfig8}(d)]. The above calculations demonstrate that the layer filter constructed by $\rm{MnBi_2Te_4}$ is capable of filtering electronic currents with specific layer-polarization. Therefore, we have theoretically proved the feasibility of fabricating the layer filter device with $\rm{MnBi_2Te_4}$.

The layer valve device is designed with the AFM $\rm{MnBi_2Te_4}$ by connecting the top-layer filter (TLF) and (BLF) from the left to the right, as sketched in Fig. \ref{sfig9}(a). A sizable surface band gap $\Delta$ of about $2M_z$=0.08 eV is induced in the device by the intrinsic magnetism, as shown in the energy spectrum [see Fig. \ref{sfig9}(b)]. Inside the gaps, a pair of linear dispersion appears both in the TLF and BLF. However, the chirality of the DW modes in the TLF and BLF on the same surface is opposite, as illustrated by the blue arrows in Fig. \ref{sfig9}(a). The layer valve device can switch the layer-polarized currents on and off. As shown in Figs. \ref{sfig9}(c) and (d), the top-layer current $J_t$, bottom-layer current $J_b$ and $J=J_t+J_b$ all vanish when Fermi energy $E_F$ is located at the DW states of both the TLF and BLF ($-1<\mu_{L}/M_z <1$ and $-1<\mu_{R}/M_z <1$). The layer valve is tuned off. By independently shifting $E_F$ out of the gap to the surface states of the TLF or BLF, a layer polarized current is generated with the layer valve switched on. Specifically, as shown in Fig. \ref{sfig9}(c), by tuning $E_F$ to the surface states of the TLF ($\mu_{L}/M_z <-1$ or $\mu_{L}/M_z >1$), $J_t$ keeps zero and $J_b$ rises to $e^2V/h$, implying that a quantized and bottom-layer polarized current is switched on. Contrarily, the current with top-layer polarization can be induced by shifting $E_F$ to the surface states of the BLF ($\mu_{R}/M_z <-1$ or $\mu_{R}/M_z >1$), as shown in Fig. \ref{sfig9}(d). From above, we demonstrate that the designed layer valve with AFM $\rm{MnBi_2Te_4}$ can realize a switch on/off of the layer-polarized current with specific layer-polarization that is manipulated by independently tuning $\mu_{L}$ of the TLF or $\mu_{R}$ or the BLF.

The performance of layer valve is also robust against weak disorders. As shown by the green dashed line in Fig. \ref{sfig9}(e), the transmission current $J$ slightly deviates from zero when $E_F$ sits both inside the surface gaps of the TLF and BLF (-1$<\mu_{L}/M_z <$1 and -1$<\mu_{R}/M_z <$1) under weak disorders $W$=0.2 eV, compared to that in the clean system ($W$=0 eV) shown by the blue solid line. However, the on-off ratio remains large since the scattering between the two DW modes of TLF and BLF is not much strengthened. Thus, the layer valve constructed with AFM $\rm{MnBi_2Te_4}$ maintains functional and robust against weak disorders. In the presence of very  strong disorders with $W$=0.5 eV (the surface gap approximates 0.08 eV), as shown by the pink dotted line in Fig. \ref{sfig9}(e), $J$ in -1$<\mu_{L}/M_z <$1 climbs up to a moderate value, implying a leakage of the transmission current. Out of the surface gap, $J$ cannot reach the quantized value $e^2V/h$ for that the tunneling electrons by the surface states of the TLF are scattered by disorders. Both aspects lower the on-off ratio of the layer valve. Whereas, we can enhance the performance of the layer valve by suppressing the scattering between the DW modes, which is realized by increasing their spatial separation $D$ as shown in Fig. \ref{sfig9}(a). When $D$ rises from $11a$ to $21a$ and $41a$, the leaked current $J$ lowers down to almost zero [see Fig. \ref{sfig9}(f)]. Meanwhile, $J$ in $\mu_{L}/M_z <-1$ or $\mu_{L}/M_z >-1$ is not much affected. The on-off ratio of the layer valve is then enhanced. We conclude that it is experimentally feasible to construct high performance layer valve device with the AFM $\rm{MnBi_2Te_4}$.

\section{Influence of the domain wall structures}
In the main text, we have assumed that the magnetic DW between the two AFM AIs in the layer filter is atomically sharp. However, in experimentally grown AFM $\mathrm{MnBi_2Te_4}$, the DW can be inherently wide \cite{magimagAI2020} which may bring non-negligible finite-size effect. In addition, there is intrinsic spin structure within the DW, across which the magnetization smoothly varies to flip the spin order in the two domains of AFM $\mathrm{MnBi_2Te_4}$ \cite{magimagAI2020,spinflop2020}. Moreover, the shapes of the DWs can also be naturally curved as observed on the surfaces of $\mathrm{MnBi_2Te_4}$ \cite{magimagAI2020,spinflop2020}. We discuss, in this section, the effects of the finite-size, the spin-flop states, and shapes of the DWs on the transport properties of the layer filter device. Because of the topological origin of the DW modes, the layer-polarized and quantized transport is supposed to be robust and insensitive to the finite-size, the shapes and the spin structures of the DWs. Thus, it is achievable to realize the layer filter device in experiment with realistic materials, such as $\mathrm{MnBi_2Te_4}$.

\subsection{The finite-size effect of the DW} 
\begin{figure}[h]
	\centering
	\includegraphics[width=15cm]{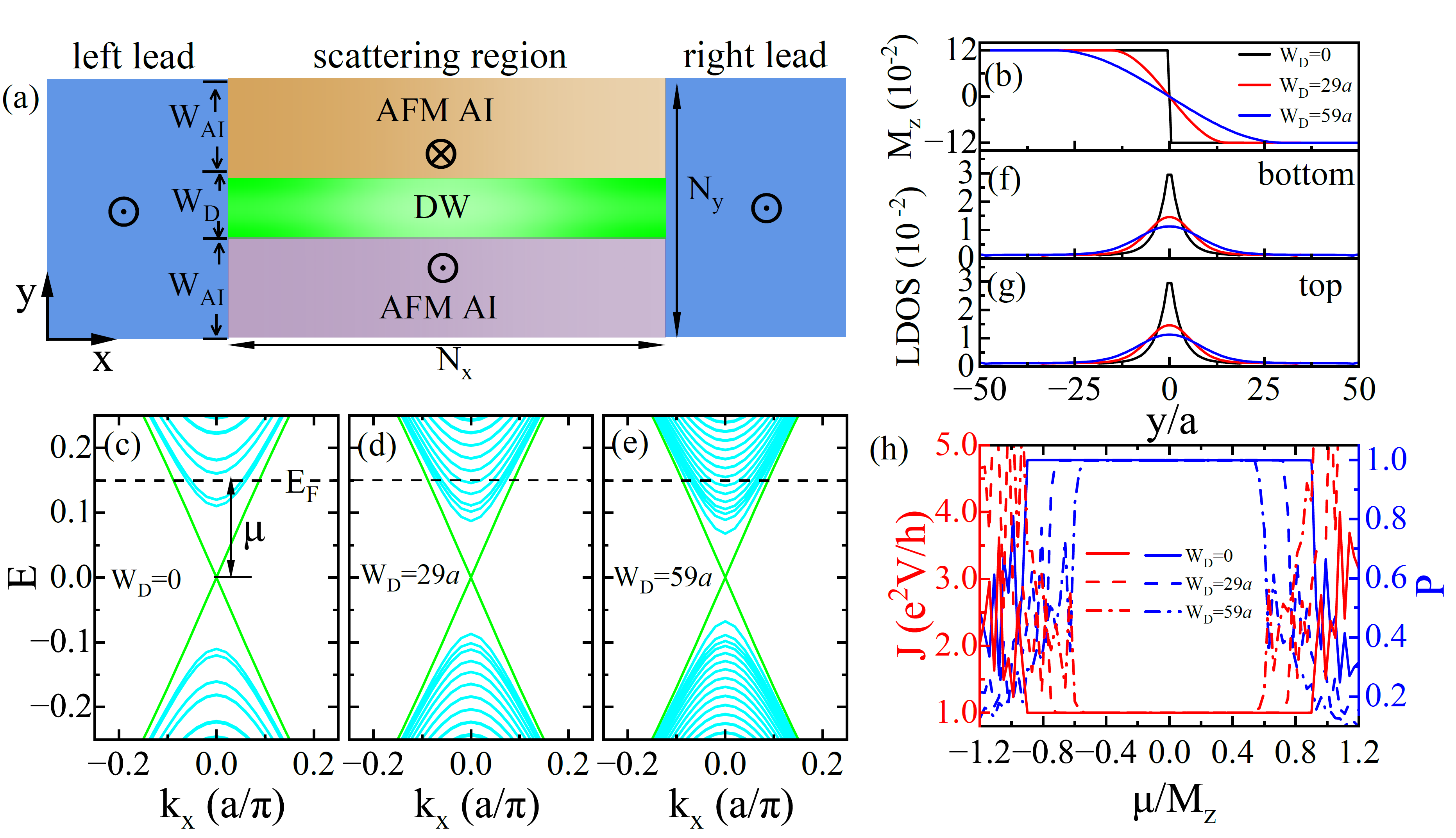}
	\caption{(color online) (a) Top-view of the layer filter with smooth DW, whose width $W_D$ is variable. The width $W_{AI}$ of the AFM AI is fixed to 19$a$. $a$=1.35 nm measures the distance between two adjacent layers. (b) The top-surface perpendicular magnetization $M_z$ of the layer filters with $W_D$=0, 29$a$ and 59$a$. (c)-(e) Energy spectrum for the layer filters with $W_D$=0, 29$a$ and 59$a$, respectively. (f)-(g) The bottom- and top-surface LDOS of the DW states for layer filters with $W_D$=0, 29$a$ and 59$a$. The energy is $E$=0.01. (h) The transmission current $J$ and its layer polarization $P$ versus chemical potential $\mu$ of the layer filters with $W_D$=0, 29$a$ and 59$a$. $\mu$ is defined as the difference between Fermi energy $E_F$ and the Dirac point of the DW modes [see (c)]. Model parameters are: $M_z$=0.12, $N_z$=10, $N_x$=100, $W_{AI}$=19a and chemical potentials of the two leads 0.2.}
	\label{sfig2}
\end{figure}

The schematic diagram of the layer filter is displayed in Fig. \ref{sfig2}(a), where the boundaries of the AFM AIs form a smooth DW with finite width $W_D$. On the top/bottom surface, the perpendicular magnetization across such a DW is described by a smooth cosine function, i.e., $\pm M_z\cos[\frac{\pi}{W_D+2a}(y+\frac{W_D}{2}+a)]$, and the magnetization is homogeneous away from the DW, as shown in Figs. \ref{sfig2}(a) and (b).
In general, the finite $W_D$ of the DW inevitably induces the trivial sub-band states on its surfaces whose gap $E_g$ is related to $W_D$ and can be approximated by $E_g=\frac{2\pi A_2}{W_D}$ if the DW has no spin structure inside ($M_z=0$ across the DW) \cite{gapkink2020}. Here, $A_2$ is the Fermi velocity of the Dirac surface states in the DW.
As $W_D$ is increased large enough, the confinement induced sub-band gap $E_g$ could decrease even lower than the intrinsic-magnetism raised surface gaps in AFM AIs (about $ 2M_z$), thus influencing the layertronic transport in the layer filter. That is to say, the dissipationless transport is possibly affected by the sub-band gap $E_g$. When taking the smooth spin structure of the DW into consideration, the sub-band gap $E^{\prime}_g$ should also exist. 

In order to illuminate the finite-size effect of $W_D$ with reduced computational resources, we first increase the magnetization $M_z$ from 0.05 to 0.12 to create a larger magnetic surface gap. Then, we fix the width $W_{AI}$ of the AFM AI, but only vary the DW width $W_D$, as show in Fig. \ref{sfig2}(a). When $W_D$ increases from 0 to 29$a$ and 59$a$, the sub-band gap of the DW states reduces [see Figs. \ref{sfig2}(c)-(e)]. Especially as shown in Fig. \ref{sfig2}(e), the sub-band gap ($E^{\prime}_g\approx 1.125 M_z$=0.135) is much smaller than the magnetic surface gap ($\approx 2M_z$=0.24), identifying the existence of the trivial 1D sub-band states in the DW. 
However, the topologically protected chiral DW modes persist in the sub-band gaps, independent of $W_D$ [see green linear lines in Figs. \ref{sfig2}(c)-(e)]. The effect of increasing $W_D$ is barely broadening the DW states, as shown by the local density of states on the top and bottom surfaces in Figs. \ref{sfig2}(f) and (g). The electronic transport currents by the DW modes preserve to be quantized ($J=e^2V/h$) and layer-polarized ($P$=+1)[see Fig. \ref{sfig2}(h)], which clarifies the robustness of the layer filter against smooth domain walls. Nevertheless, the chemical potential $\mu$-window where the layer filter works shrinks from the magnetic gap ($\approx 2M_z$) to the sub-band gap $E^{'}_g$ due to the finite-size effect of the DW, as shown in Fig. \ref{sfig2}(h).  We see above that the layertronic and quantized transport occurs in the sub-band gap $E^{\prime}_g$ of the DWs and thus determined by $W_D$. The larger $W_D$ is, the narrower $\mu$-window the layer filter works.

We further give an estimation of the critical $W_D$ where the electronic transport by the DW modes can be distinguished in realistic material $\mathrm{MnBi_2Te_4}$. Recall that the confinement-induced sub-band gap of the DW without spin structure is approximated by $E_g=\frac{2\pi A_2}{W_D}$. Taking the parameters of the effective model in Fig. \ref{sfig2}(e), we have $E_g\approx 0.488 M_z$=0.0586. From that $E_g<E^{\prime}_g$, it is reasonably believed that the smooth domain wall spin structure renormalizes the realistic DW width by a factor $\alpha$, i.e., $E^{\prime}_g=\frac{2\pi A_2}{\alpha W_D}$. Then we obtain $\alpha=\frac{E_g}{E^{\prime}_g}\approx$ 0.43. The $\alpha$ is expected to be insensitive to the parameters if the spin structure insider the DW is smooth. Therefore it can be used as a constant to estimate the DW width with realistic material parameters.
From the transport experiment of HgTe/CdTe\cite{QSHHgTe2006,QSHHgTe2007}, it is proved that the topological transport can be detected as long as the bulk gap remains larger than about $\Delta$=10 meV. Therefore for realistic $\rm{MnBi_2Te_4}$, the DW width $W_D$ is required to be smaller than $\frac{2\pi A_2}{\alpha \Delta}\approx$ 468 nm ($A_2\approx$ 3.2 eV$\cdot${\rm{\AA}} is proportional to the Fermi velocity of the Dirac surface states for $\rm{MnBi_2Te_4}$). We ensure that in such a condition, the layer filter keeps working and the layertronic transport can be detected. Note that the observed DW width in experiment is about a few hundred nanometers on the surfaces of $\rm{MnBi_2Te_4}$ \cite{magimagAI2020,spinflop2020}, which quantitatively meets the requirement and makes $\rm{MnBi_2Te_4}$ an ideal platform to construct layertronic devices.

The above estimation is based on the case that $\mathrm{MnBi_2Te_4}$ is in clean condition. The disorder should exist in experimental samples. Meanwhile, the number of the trivial sub-bands around the DWs is small since they are spaced by $E^{\prime}_g/2$ in energy regime. The 1D nature of the trivial states further makes them easier to be localized by disorders, expanding the $\mu$-window where DW-mode transport occurs and the layer filter works. Therefore, quantized and layer-polarized transport can be observed in $\rm{MnBi_2Te_4}$ with even wider DWs than the estimated 468 nm.

\subsection{The influence of the non-uniformity and different spin-flop states inside the DW}
The studies on the surface spin-flop transitions of the AFM $\rm{MnBi_2Te_4}$ have demonstrated that the magnetic order within the DW is a canted AFM order \cite{magimagAI2020,spinflop2020}. Besides the perpendicular component $M_z$, other two magnetization components ($M_x$ and $M_y$) that are parallel to the terminated surface are possibly finite inside the DWs.
\begin{figure}[h]
	\centering
	\includegraphics[width=14cm]{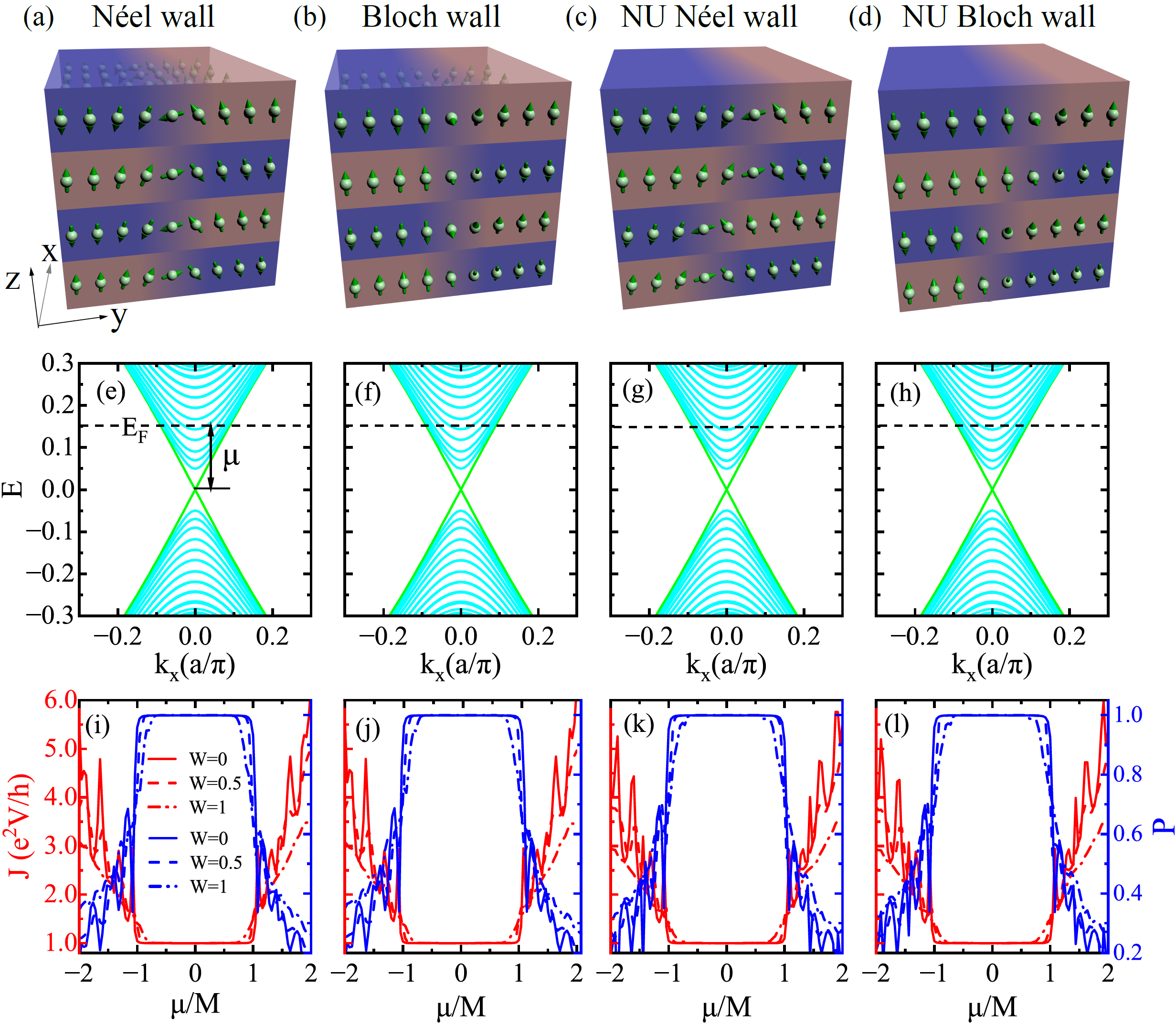}
	\caption{(color online) (a)-(d) Schematic diagrams of the Néel, Bloch, nonuniform (NU) Néel and NU Bloch magnetic DWs (shadow regions). Green arrows with spheres illustrate the spin-flop on each layers. For Néel (Bloch) wall, the spin flops along $y$ ($x$) direction. For NU Néel and NU Bloch walls, the magnetic DW on each layers is displaced randomly, as shown by the shadow regions. (e)-(h) Corresponding energy spectrum of the layer filters with the four kinds of DW structures. (i)-(l) The corresponding current $J$ and layer polarization $P$ versus chemical potentials $\mu/M$, with ($W$=0.5, 1) or without ($W$=0) disorders. Model parameters are: $M$=0.05, $N_z$=10, $N_y$=60, $N_x$=80, $W_D$=19$a$ and $\mu$ of the two leads is $4M=0.2$. For the NU Néel and NU Bloch walls, the displacement of the magnetic DW on each layer is randomly chosen to be (5$a$ 7$a$ -6$a$ 7$a$ 2$a$ -7$a$ -4$a$ 1$a$ 8$a$ 8$a$).}
	\label{sfig3}
\end{figure}
To explore the effects of the DWs with realistic spin structures, some complementary calculations are performed and given in the Fig. \ref{sfig3}. 
Two energetically favorable DW spin structures, the N\'{e}el wall and Bloch wall \cite{neelwall2013,neelbloch2013}, are considered. These two DW configurations can be described by a magnetization vector $\mathbf{M}(y)=(M_x, M_y, M_z) = M(\sin\theta\cos\phi, \sin\theta\sin\phi, \cos\theta)$, where $M$ is a constant magnetization magnitude originating from magnetic doping \cite{QTonDW2017,configDW2018}. The azimuthal angle $\phi$ defines the type of DW. $\phi=0$ and $\phi=\frac{\pi}{2}$ represent the N\'{e}el wall and Bloch wall respectively, as shown by the schematic diagrams in Figs. \ref{sfig3}(a)-(b). And the azimuthal angle $\theta$ is a function of $y$ with $\cos\theta(y)=-\tanh\frac{y}{W_D}$. The AFM magnetization is homogeneous away from the DW. We calculate the energy spectrum of the layer filters with the N\'{e}el and Bloch DWs [see Figs. \ref{sfig3}(e)-(f)]. The DW modes emerge in the surface band gaps as expected, which originate from the topology difference across the DWs. The transmission currents through the filter carried by the DW states are quantized ($J=e^2V/h$) and layer-polarized ($P=+1$) [see red and blue solid lines in Figs. \ref{sfig3}(i)-(j)], which evidently confirms that layer filters with the two DW structures are desirably functional. When disorders are involved in, the transport properties ($J=e^2V/h$ and $P=+1$) persist owning to the topology and robustness of the DW states against disorders [see the red, blue dashed and dot-dashed lines in Figs. \ref{sfig3}(i)-(j)]. 

Furthermore, we have also studied the non-uniformity of the DWs across the sample from the top to bottom layers. The magnetic DW on each layer is displaced randomly, as shown in Figs. \ref{sfig3}(c)-(d). We plot the energy spectrum of the layer filters with nonuniform Néel and Bloch walls in Figs. \ref{sfig3}(g)-(h). The non-uniformity brings little impacts and the DW modes in the surface gaps still exist. The characteristics of the electronic transport by the DW states with and without disorders are similar as that of the filters with uniform Néel and Bloch DWs [see Figs. \ref{sfig3}(i)-(l)]. Thus, we conclude that the layer filter device can perform as well as designed, regardless of the spin structures, or the spin-flop states, inside the DWs between the AFM AIs.

\subsection{The influence of the curved DW}
\begin{figure}[h]
	\centering
	\includegraphics[width=12.5cm]{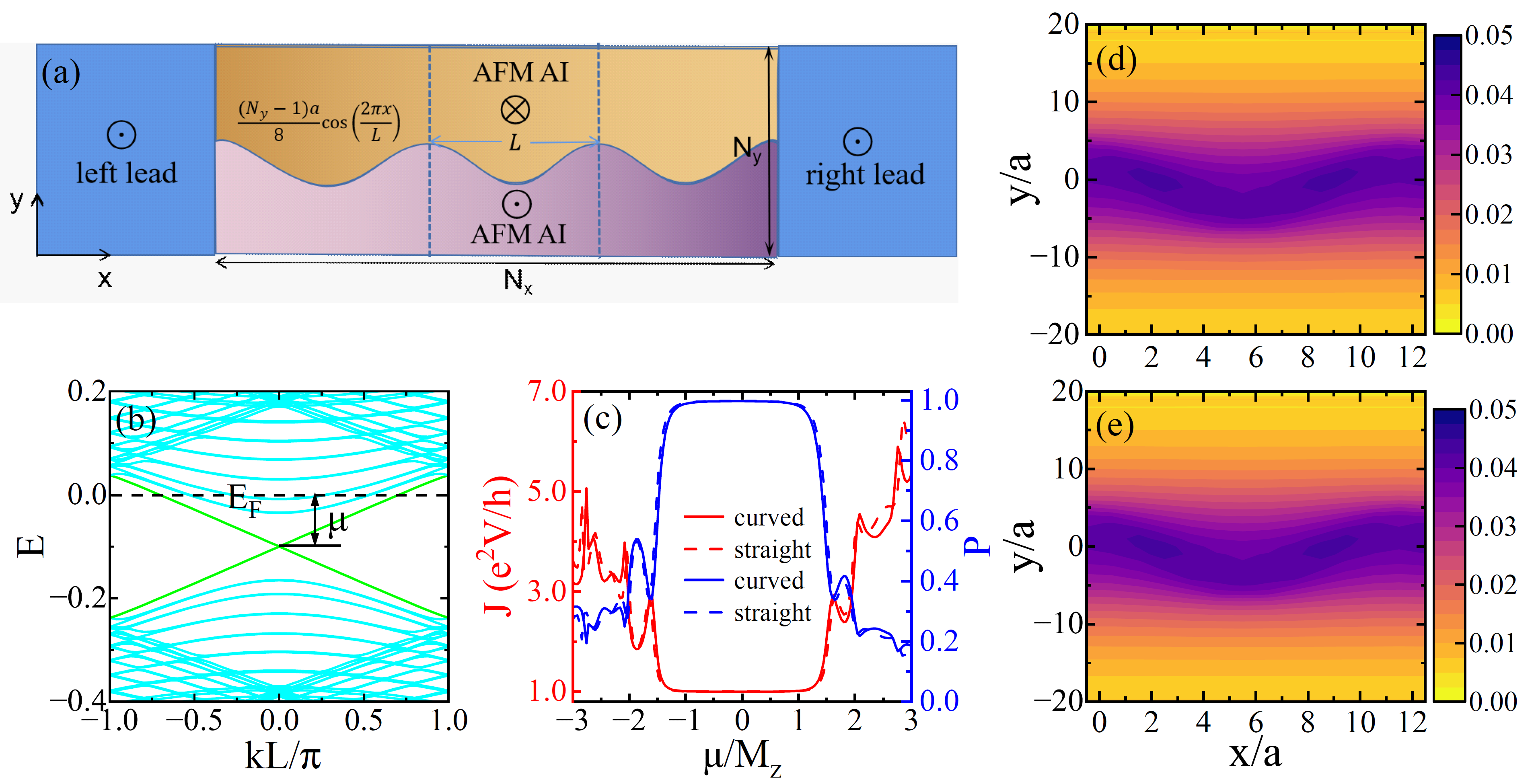}
	\caption{(color online) (a) Top-view of the layer filter with curved DW described by $y=\frac{(N_y-1)a}{8}\cos(\frac{2\pi x}{L})$, where $L$ is the period. The vertices represent the surface magnetization directions (in and out) of each part. (b) Folded energy spectrum of the layer filter. (c) The transmission current $J$ and its layer polarization $P$ of the layer filters with curved DW (solid lines) and straight DW (dashed lines), versus chemical potential $\mu/M_z$. The left/right lead is the source/drain terminal. (d) Energy-resolved ($E=0.01$) top and (e) bottom surface LDOS of the DW states inside the surface band gap. Model parameters are: $M_z=0.05$, $N_z=10$, $N_y=40$, $N_x=36$, $L=12a$ and $\mu$ of the leads is $4M_z=0.2$.}
	\label{sfig4}
\end{figure}

The curved shape of the magnetic DWs is intrinsic in experimentally fabricated $\rm{MnBi_2Te_4}$ \cite{magimagAI2020}.
We here simulate the curved domain wall by a cosine function $y=\frac{(N_y-1)a}{8}\cos(\frac{2\pi x}{L})$, as shown in the central part of Fig. \ref{sfig4}(a). The topological DW modes are expected to emerge at the interface regardless of the shapes of the DWs. Their existence is manifested by the energy spectrum where two linear dispersion crosses over the surface band gaps, as shown Fig. \ref{sfig4}(b). 
We numerically plot the LDOS on the top and bottom surfaces of the chiral states inside the surface gap, as given in Figs. \ref{sfig4}(d)-(e). The LDOS peaks nearby the curved interface, which further verifies that the chiral states are well-confined DW modes. The layer filter with curved DW shares similar electronic transport properties as the layer filter with straight DW by comparing their transmission current $J$ and layer-polarization $P$, as shown in Fig, \ref{sfig4}(c). Therefore, the layer filter with curved DWs can indeed filter quantized ($J=e^2V/h$) and fully layer-polarized ($P=+1$) currents. In conclusion, the layer filters maintain robust against curved DW structures.

\section{Local current distributions in the layer valve}
We have seen in the main text that the layer-polarized current can be switched on or off in the layer valve device,
\begin{figure}[h]
	\centering
	\includegraphics[width=10cm]{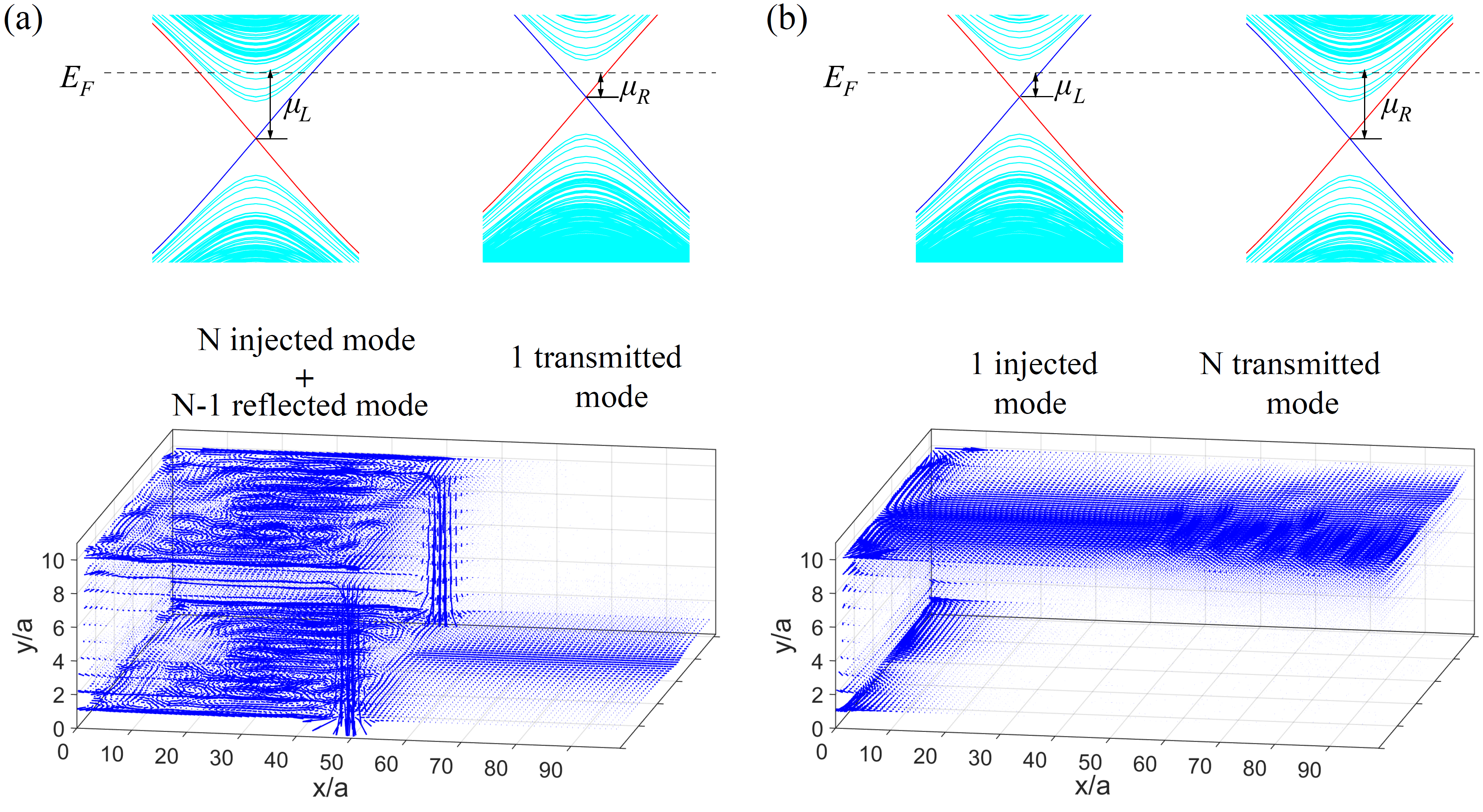}
	\caption{(color online) (a) Top: Schematic energy spectrum for the TLF and BLF in the layer valve, with Fermi energy $E_F$ marked by the dashed line. Chemical potential $\mu_{L}$ ($\mu_{R}$) is the difference between $E_F$ and the Dirac point of the DW modes in the TLF (BLF). Bottom: local current distribution of the bottom-layer polarized current in the layer valve, with $\mu_{R}/M_z=0.4$ and $\mu_{L}/M_z=4.1$. (b) Top: Schematic energy spectrum for the TLF and BLF. Bottom: local current distribution of the top-layer polarized current, with $\mu_{L}/M_z=0.4$ and $\mu_{R}/M_z=4.1$. Other model parameters are $M_z=0.05$, $N_z=10$, $N_y=N^L_x=N^R_x=50$, $D=11 a$ and $\mu$ of the terminals is $5M_z=0.25$.}
	\label{sfig5}
\end{figure} 
and its layer polarization is manipulated by independently controlling the Fermi level to the surface states of the TLF or of the BLF. 
We here display the local current distributions of the layer-polarized transmission currents when the valve is tuned on. 
For the first kind of ``on'' status, as shown in the bottom panel of Fig. \ref{sfig5}(a), the current flows into the right terminal totally from the bottom surface with $P=-$1. The majority of electrons flowing to the interface ($x/a=50$) between the TLF and BLF are reflected, because the number of surface state transverse modes in the TLF is significantly larger than that of the DW mode in the BLF [see top panel of Fig. \ref{sfig5}(a)].
The local current distribution is determined by the interference of the $N$ injected and $N-1$ reflected current modes. The situation differs in the second case. As shown in the bottom panel of Fig. \ref{sfig5}(b), the current flows to the right terminal entirely from the top surface ($P=+1$).  The electrons carried by the upper DW mode of the TLF transmit to the top surface of the BLF without reflection, since only one injected mode in the TLF carries electrons to transport to the large $N$ surface state transmitted modes in the BLF [see the top panel of Fig. \ref{sfig5}(b)]. The local current distribution is determined by the perfect transmission mode.

The surface transport for the two ``on'' status of the layer valve is apparently different, as shown in Fig. \ref{sfig5}. In spit of the mismatch between the number of transverse modes in the TLF and BLF discussed above, another aspect is the applied bias. Since a voltage drop is imposed between the leads to drive electrons to flow from left to right in both cases, there should be no symmetry connecting the surface transport in Figs. \ref{sfig5}(a) and (b).


\section{Reversal of the layer polarization of the transmission currents in the layer valve}
\begin{figure}[h]
\centering
\includegraphics[width=14cm]{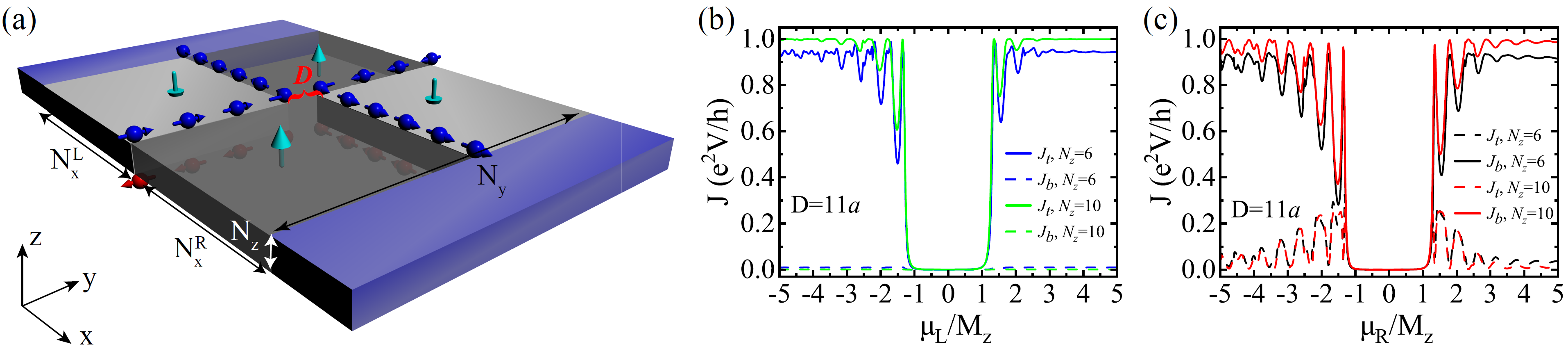}
\caption{(color online) (a) Sketch of the layer valve consisting of BLF and TLF that are connected in sequence from left to right. (b) Layer-resolved transmission currents $J_t$ and $J_b$ versus chemical potential $\mu_{L}/M_z$ of the left part (BLF) of the valve, with fixed chemical potential $\mu_{R}/M_z=0.4$ of the right part (TLF). (c) $J_t$ and $J_b$ versus $\mu_{R}/M_z$, with fixed $\mu_{L}/M_z=0.4$. Model parameters are $M_z=0.05$, $N_y=N^L_x=N^R_x=50$, $D=11 a$ and $\mu$ of the leads is $5M_z=0.25$.}
\label{sfig6}
\end{figure}

In this section, we show that by exchanging the positions of the TLF and BLF in the layer valve or by inverting the source and drain terminals in the main text, we are able to reverse and thus control the layer polarization of the transmission current when layer valve is tuned on. The newly structured layer valve is presented in Fig. \ref{sfig6}(a). The chirality of all DW modes reverses on each part, compared to that in the main text [see Fig. 3(a)]. When either chemical potential $\mu_{L}$ of the left part (BLF) or $\mu_{R}$ of the right part (TLF) is located at the surface states to switch on the valve, the layer polarization of the transmission current is reversed as well, contrary to that in the main text [see Figs. 3(b)-(c)]. Specifically, as shown in Fig. \ref{sfig6}(b), when $\mu_{L}$ is shifted from the DW states in the gap (-1$<\mu_{L}/M_z<$1) to the surface states ($\mu_{L}/M_z<$-1 or $\mu_{L}/M_z>$1), the top layer transmission current $J_t$ increases significantly from zero to about $e^2V/h$ while bottom transmission current $J_b$ keeps vanishing. A top-layer polarized current is generated [bottom-layer polarized current in the main text as shown in Fig. 3(b)]. On the other hand, when tuning $\mu_{R}$, $J_b$ arises from 0 to $e^2V/h$ with vanishing $J_t$, as seen in Fig. \ref{sfig6}(c). A bottom-layer polarized current is switched on [top-layer polarized current in the main text as shown in Fig. 3(c)]. The layer valve proposed here can generate layer polarized current whose layer polarization is opposite to that in the main text when shifting either $\mu_{L}$ of the left part or $\mu_{R}$ of the right part to the surface states. Note that the exchange of the TLF and BLF might be achieved using electromagnetic field to reverse the AMF order in each AI domain through the $\mathbf{E}\cdot \mathbf{B}$ term \cite{LHE2021}.

\section{Layer valve with higher on-off ratio}
\begin{figure}[h]
	\centering
	\includegraphics[width=14cm]{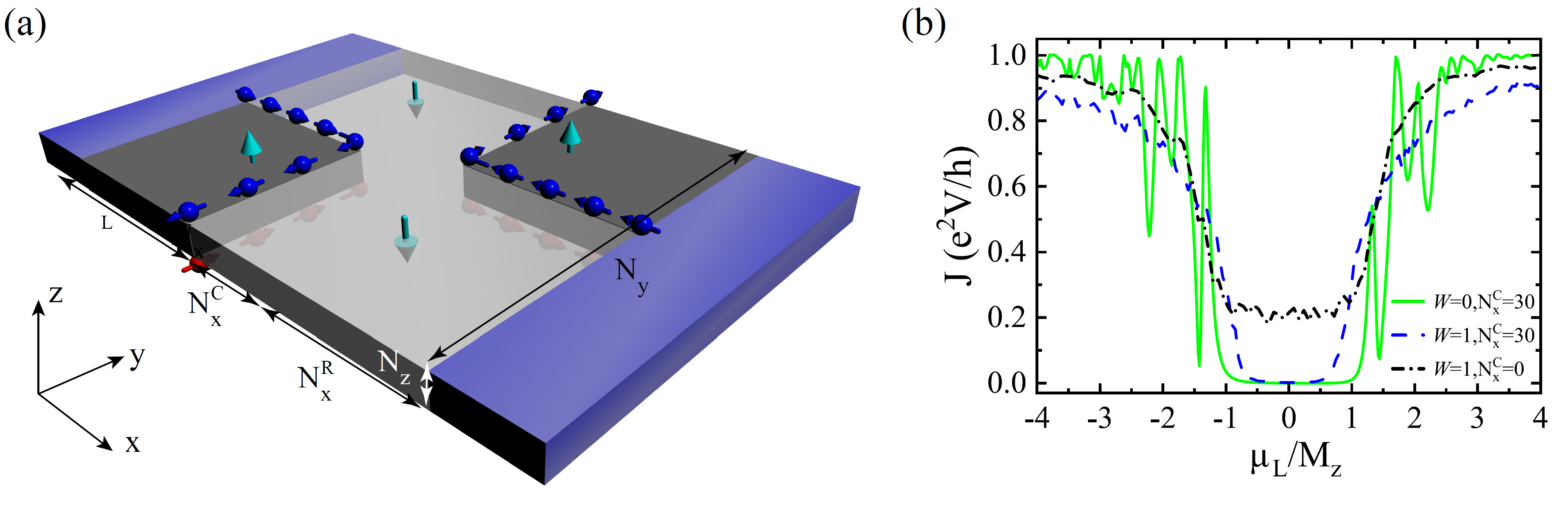}
	\caption{(color online) (a) Schematic diagram of the layer valve consisting of TLF, AFM AI and BLF that are connected in sequence from left to right. (b) Transmission current $J$ of this new valve (blue dashed line $N^C_x=30$) and of the valve in the main text (black dot-dashed line $N^C_x=0$) in the presence of disorders ($W=1$), versus $\mu_{L}/M_z$ of the TLF and AFM AI, with fixed $\mu_{R}/M_z=0.4$ of the BLF. Green solid line is the $J$ of this new valve in absence of disorders, plotted for comparison. $\mu$ of the central AFM AI is set the same as that of the TLF. Model parameters are: $M_z=0.05$, $N_z=10$, $N_y=N^L_x=N^C_x=N^R_x=30$ and $\mu$ of the leads is $5M_z=0.25$.}
	\label{sfig7}
\end{figure}

We propose that by inserting antiferromagnetic AI domain between the TLF and BLF of the layer valve in the main text [see Fig. 3(a)], the on-off ratio of the valve can be further enhanced in the presence of disorders, due to a larger separation and reduction of the scattering between the DW states in the TLF and BLF. 
This newly constructed layer valve is shown in Fig. \ref{sfig7}(a). A central part of antiferromagnetic AI domain is inserted to separate the DW states further in $x$ direction. As a consequence, the scattering between DW states in TLF and BLF assisted by disorders, is lowered and a higher on-off ratio is achieved compared to that in the main text. We plot together the transmission current $J$ versus $\mu_{L}/M_z$ of this new valve and of the layer valve in the main text in the presence of disorders [see blue dashed and black dot-dashed lines in Fig. \ref{sfig7}(b)]. When $\mu_{L}$ sits in the gap of the surface states (-1$<\mu_{L}/M_z<$1), $J$ of this new valve is smaller than that of the valve in the main text, indicating that this new layer valve possesses a higher on-off ratio in the presence of disorders.

\section{Realizing layer filter and layer valve devices in a gate-controlled way}
\begin{figure}[h]
	\centering
	\includegraphics[width=14cm]{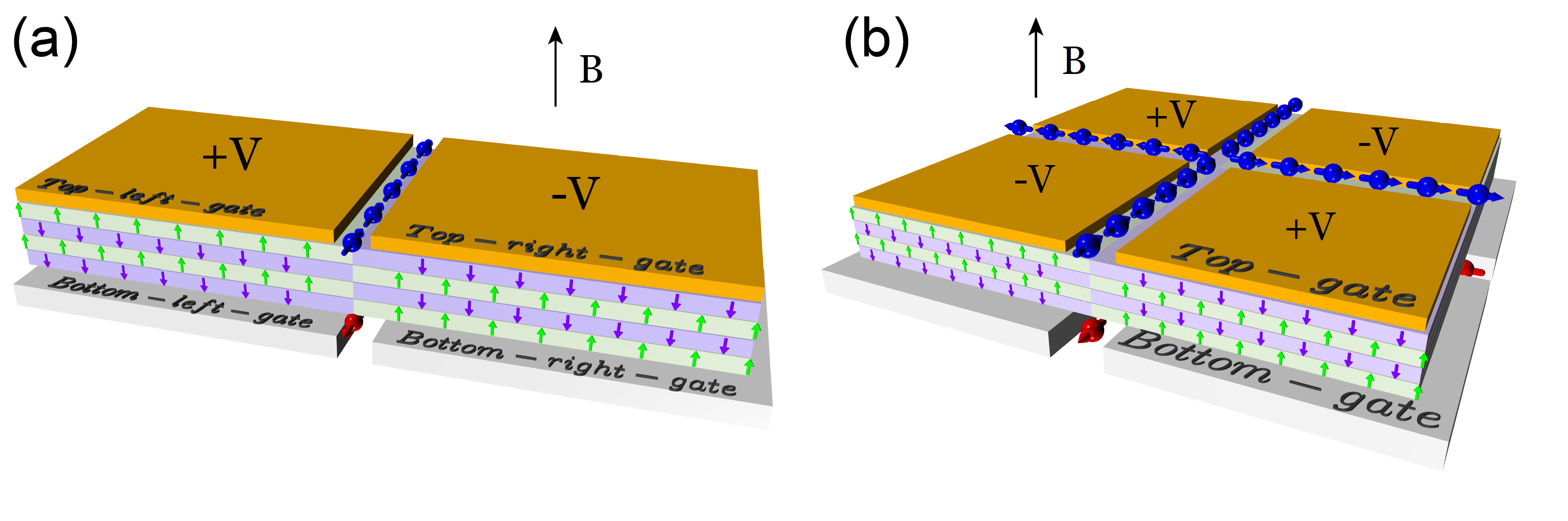}
	\caption{(color online) Schematic of gate-controlled layer filter and layer valve devices in AFM $\rm{MnBi_2Te_4}$. (a) The formation of a single domain wall to realize the layer filter can be achieved through a pair of split gates. Due to the magnetoelectric coupling of AIs, opposite gate voltages ($V $ or $-V$) on the gates induce opposite perpendicular electric field $\textbf E$, giving rise to a heterostructure of the two types of AFM orders under a perpendicular magnetic field $\textbf B$. (b) The layer valve device can be similarly constructed by two pairs of split gates. Gate voltages are applied alternatively.}
	\label{sfiggate}
\end{figure}

In real experimental conditions, it may be difficult to realize the layertronic devices from naturally formed domain structures, especially the layer valve device. Fortunately, $\rm{MnBi_2Te_4}$ is Van der Waals stacked material, of which the fabrication technique is based on mechanical exfoliation. Its fabrication technique is very similar to the bilayer graphene. Therefore, motivated by the technical route of realizing the valleytronic devices in bilayer graphene \cite{Topovalley2015,GateBLG2016,valleyval2018,GateBLG2020,Ivar2008}, we propose that the layer filter and layer valve devices can similarly be realized in a gate-controlled way. The key ingredient is that in the presence of a perpendicular magnetic field $\mathbf{B}$, the AFM order of $\rm{MnBi_2Te_4}$ can be reversed by applying perpendicular electric field $\mathbf{E}$ due to the uique magnetoelectric coupling (i.e., the $\theta \mathbf{E}\cdot\mathbf{B}$) of the axion insulator. Such a process has already been realized experimentally \cite{LHE2021}. Similar to what has been realized in bilayer graphene, the layer filter can be achieved by a pair of split gates on which opposite gate voltages are applied in the presence of a perpendicular magnetic field as shown in Fig. \ref{sfiggate} (a). Due to the magnetoelectric coupling $\theta \mathbf{E}\cdot\mathbf{B}$, the domain of the perpendicular $\mathbf{E}$ gives rise to the domain of AFM order. Further, the layer valve device can be achieved by fabricating two pairs of split gates [see Fig. \ref{sfiggate} (b)]. Moreover, the bulk gap of the valley valve devices is about 86 meV \cite{valleyval2018}, while the surface gap of $\rm{MnBi_2Te_4}$ is about 50-100 meV. They are of the similar order. Therefore, the topologically nontrivial transports should emerge at similar length scale of the devices.

There are some advantages in realizing layertronic deivces in $\rm{MnBi_2Te_4}$ than realizing valleytronic devices in bilayer graphene. Firstly, the magnetization reversal can be achieved by applying an impulse voltage. Therefore, one convenience in $\rm{MnBi_2Te_4}$ is that the gate voltage need not to be applied continuously. Specifically, the gate voltages applied on bilayer graphene should be fine-tuned such that the bulk gap of oppositely-gated domains are always aligned to ensure the global insulating bulk. But in $\rm{MnBi_2Te_4}$, such a difficulty is naturally circumvented because we only need the impulse voltage to prepare the initial domain configuration. Secondly, during the repeatable switching process as shown in \cite{LHE2021}, the maximum (around +0.5 $k\Omega$) and the minimum values (around -0.9 $k\Omega$) of $R_{yx}$ do not show any decay feature, which indicates that the reversal of the AFM order is of high-efficiency and causes little magnetization loss. Therefore, making high quality layertronic devices in $\rm{MnBi_2Te_4}$ is achievable. Thirdly, the topological kink states in valleytronics are not robust to disorders. This is because the chiral domain wall modes from $K$ and $K^\prime$ valleys are not spatially separated and can be backscattered by short range disorders. Therefore, quantized transport is hard to observe. In contrast, the chiral domain wall modes in $\rm{MnBi_2Te_4}$ are spatially separated (located on the top and bottom surfaces). Backscattering between these domain wall modes requires stronger disorder. Therefore, the quantized transport properties make the dissipationless layertronics easier to be characterized. 

\section{Some details in the numerical calculations}
This section is devoted to clarify some of the details in our numerical calculations in the main text. \textbf{1)} The NEGF method is applied for the calculation of local current distribution. The current flowing from site $i$ to its neighboring site $j$ can be calculated by \cite{numHgTe2009}
\begin{equation}
    J_{i\rightarrow j}=\frac{2e^2}{h} {\rm Im} \{ \sum_{\alpha,\beta} H_{i\alpha,j\beta} \left[G^r\Gamma^LG^a\right]_{j\beta,i\alpha} \} (V_L-V_R),
\end{equation}
where $H_{i\alpha,j\beta}$ is the hopping of electrons from orbit $\alpha$ at $i$ to orbit $\beta$ at $j$, $G^r$ ($G^a$) denotes the retarded (advanced) Green's functions of the central region, $\Gamma^L$ represents the linewidth function of the left terminal, and $V=V_L-V_R$ gives the external bias. The layer-resolved transmission currents $J_t$, $J_b$, $J=J_t+J_b$ and layer polarization $P$ are evaluated by summing $J_{i\rightarrow j}$ on a specific layer perpendicular to $x$ direction that is close to the drain terminals, in order to explore the effects of layertronic devices on the unpolarized current injected from source terminals.
\textbf{2)} Due to the restriction of the theoretical 3D topological insulator model, the Dirac fermions on the side surface are inevitably involved into the electronic transport. Experimentally, the contribution of the side surface states to the transport should not count, for that the size of the experimentally fabricated $\mathrm{MnBi_2Te_4}$ is super large [e.g., see Fig. 1 in the main text. $N_x$ and $N_y$ are of the order $\mu m$, while $N_z$ is of the order $nm$]. Thus in our calculations, we incorporate a $y$-direction magnetization term $M_2(y) s_2\otimes \sigma_0$ in $H_{\rm mag}$ to gap out the side surface states. $M_2(y)=\pm M_y=\pm 0.3$ is the perpendicular magnetization on the side surfaces. \textbf{3)} We have shown the local current distributions for the layer filters and layer reversers in the main text. However, these configurations are plotted in a simplified way that the local currents whose magnitudes are smaller than a critical value, like 0.01 $e^2V/h$, are omitted for the sake of clarity. \textbf{4)} Disorder effect is also considered in the main text. The calculations for $J$ and $P$ are all averaged over 200 configurations of disorders. \textbf{5)} For the experimental scheme to verify the layer polarization reverse in the layer reverser device, we propose to attach two metallic contacts (terminals 3 and 4) at the DWs on the top surface. The density of states of the two metallic terminals are approximated by wide band approximation, and we choose the self-energy contributed by the two contacts as $\Sigma^r=-1.5i$ in our numerical calculations.

\bibliography{ref2}










